\documentclass[usenatbib]{mn2e}

\usepackage{graphicx,psfig,fancyhdr,natbib,subfigure}
\usepackage{epsfig, psfig, epsf}
\usepackage{amsmath, cancel}
\usepackage{amssymb}
\usepackage{lscape}

 \def\apjl{ApJ~Lett.} \def\apj{ApJ}
\def\apjs{ApJS} \def\aj{AJ} \def\mnras{MNRAS}
\def\prd{Phys.~Rev.~D}

 \def\aap{Astron.~\&~Astrophys.} \def\araa{ARA\&A}
\def\procspie{Proceedings~of~the~SPIE}

\def \Omlam {\Omega_{\Lambda}}
\def \Omm {\Omega_{\rm m}}
\def \ho {H_0}

\def \kms {{\rm ~km~s}^{-1}}
\def \kmsmpc {{\rm ~km~s}^{-1}~{\rm Mpc}^{-1}}
\def \hmpc{~\;h^{-1}~{\rm Mpc}}

\def \xis{\xi(s)}

\def \gsim { \lower .75ex \hbox{$\sim$} \llap{\raise .27ex \hbox{$>$}} }
\def \lsim { \lower .75ex \hbox{$\sim$} \llap{\raise .27ex \hbox{$<$}} }

\def \wp {w_{p}(\sigma)}

\def \ATLAS {VST-AA$\Omega$ {\it ATLAS} }

\begin{document}

\title[LRG Clustering at $z \simeq 0.7$.]
{Luminous Red Galaxy Clustering at $z\simeq0.7$ - First Results using AAOmega}
\author[N.P. Ross et al.]
{Nicholas P. Ross\thanks{email: Nicholas.Ross@durham.ac.uk}$^{1,2}$, 
T. Shanks$^1$, 
Russell D. Cannon$^3$,
D.A. Wake$^1$, 
R.~G. Sharp$^{3}$,  
\newauthor S.~M. Croom$^{4}$ and John A. Peacock$^{5}$\\
\\
$^1$Physics Department, Durham University, South Road, Durham, DH1 3LE, UK\\
$^2$Department of Astronomy and Astrophysics, The Pennsylvania State University,
525 Davey Laboratory, University Park, PA 16802, U.S.A.\\
$^3$Anglo-Australian Observatory, PO Box 296, Epping, NSW 1710, Australia\\ 
$^4$School of Physics, University of Sydney, Sydney, NSW 2006, Australia\\
$^5$SUPA, Institute for Astronomy, University of Edinburgh, Royal Observatory, 
    Blackford Hill, Edinburgh EH9 3HJ
}
\date{\today}
\maketitle

\begin{abstract}
We report on the AAT-AAOmega LRG Pilot observing run to establish the
feasibility of a large spectroscopic survey using the new AAOmega
instrument.  We have selected Luminous Red Galaxies (LRGs) using
single epoch SDSS $riz$-photometry to $i<20.5$ and $z<20.2$. We have
observed in 3 fields including the COSMOS field and the COMBO-17 S11
field,  obtaining a sample of $\sim$600 redshift $z\gtrsim0.5$ LRGs.
Exposure times varied from 1 - 4 hours to determine the minimum
exposure for AAOmega to make an essentially complete LRG redshift
survey in average conditions. We show that LRG redshifts to $i<20.5$
can measured in $\approx$1.5hr exposures and present comparisons with
2SLAQ and COMBO-17 (photo-)redshifts. Crucially, the $riz$ selection
coupled with the 3-4$\times$ improved AAOmega throughput is shown to
extend the LRG mean redshift from $z$=0.55 for 2SLAQ to $z=0.681\pm
0.005$ for $riz$-selected LRGs. This extended range is vital for
maximising the S/N for the detection of the baryon acoustic
oscillations (BAOs). Furthermore, we show that the amplitude of LRG
clustering is $s_0 = 9.9\pm0.7\hmpc$, as high as that seen in the
2SLAQ LRG Survey. Consistent results for this clustering amplitude are
found from the projected and semi-projected correlation
functions. This high amplitude is consistent with a long-lived
population whose bias evolves as predicted by a simple ``high-peaks''
model. We conclude that a redshift survey of \hbox{360 000} LRGs over
3000 deg$^{2}$,  with an effective volume some $4\times$ bigger than
previously used to detect BAO with LRGs, is possible with AAOmega in
170 nights.
\end{abstract}


\section{Introduction}

Large-scale structure (LSS) studies are one road into investigating
``Dark Energy'' (DE) and its potential evolution
\citep[e.g.][]{BlakeGlazebrook03, Seo03, Seo05, Seo07, Angulo08}.
This has been powerfully demonstrated by recent results from the
Luminous Red Galaxy (LRG) Sloan Digital Sky Survey (SDSS),
\citep[e.g.][]{Eisenstein05, Tegmark06, Percival07a, Percival07b}  and
indeed the 2dFGRS \citep{Cole05}.  Luminous Red Galaxies (LRGs) are
predominantly massive early-type galaxies and are  intrinsically
luminous ($\gtrsim 3L^{*}$) \citep{Eisenstein03, Loh06, Wake06}.  They
are strongly biased objects, having values of $b\sim2$,
\citep{Padmanabhan07}  where $b$ is the linear bias and relates, in
the linear regime, the underlying mass density distribution to that of
the luminous tracers via  $\delta_{g} = b \, \delta_{m}$.  As such and
coupled to their very clean and efficient selection,  LRGs are
excellent tracers of large-scale structure  and can be used as
cosmological probes. \citet{Eisenstein05}, \citet{Tegmark06},
\citet{Hutsi06}, \citet{Percival07a} and \citet{Percival07b} use
positions and spectroscopic redshifts from the SDSS LRG Survey in
order to accurately measure the correlation function and the Power
Spectrum. Specifically, a detection of the baryon acoustic
oscillations (BAOs) in the galaxy distribution is made.  BAOs in the
galaxy distribution are caused by sound waves propagating through the
baryon-photon plasma in the early ($z > 1100$) Universe. At
recombination, these sound waves are ``frozen'' into the distribution
of matter at a preferred scale \citep[see e.g.][for further BAO
details]{Eisenstein98, Meiksin99, Yamamoto06, Eisenstein07}. With
measurements of the BAOs now starting to appear feasible, there is a
push to carry out large galaxy surveys at higher redshift, with the
primary goal of tracking the evolution of dark energy and the related
equation of state parameter, $w_{\rm DE}(z)$, over cosmic time. As
such, several new galaxy redshift surveys have been proposed.
%
\begin{table*}
  \begin{center}
    \setlength{\tabcolsep}{4pt}
    \begin{tabular}{lrrcccccccccccc}
      \hline
      \hline
      Field Name   & R.A. (J2000)   & Dec (J2000)       & No. of exposures & \multicolumn{5}{c}{Average seeing($''$)} & &   \multicolumn{5}{c}{Average airmass} \\
      \hline  
      COSMOS       & 10h 00m 28.6s  & \,02d 12m 21.0s   & 0+7+0+6+0 & --  & 2.0 & --  & 3.0 & --                  &     & --   & 1.39 & --   & 1.27 & --   \\
      COMBO-17 S11 & 11h 42m 58.0s  & $-$01d 42m 50.0s  & 2+6+4+0+9 & 2.0 & 1.8 & 1.7 & --  & 1.9                 &     & 1.15 & 1.19 & 1.21 & --   & 1.19 \\ 
      2SLAQ d05    & 13h 21m 36.0s  & $-$00d 12m 35.0s  & 8+0+0+5+0 & 1.9 & --  & --  & 1.6 & --                  &     & 1.22 &  --  &  --  & 1.19 & --   \\ 
      \hline
      \hline
      \label{tab:The_AAOmega_fields}
    \end{tabular}
\caption[Details of the 3 AAOmega LRG Pilot fields]
	{The 3 AAOmega LRG Pilot fields.   
         The fourth column gives the number of 1200 second exposures 
         on the 5 consecutive nights of the pilot run, 
         03 March 2006 through 07 March 2006. Note that the 9 exposures taken
	 in the S11 field on the night of 07 March 2006 targeted objects
	 which had a $z$-band magnitude selection of $19.5 < z <20.2$. }
  \end{center}
\end{table*}
One possibility is to use the AAOmega spectrograph at the AAT to make
a spectroscopic redshift survey of high redshift LRGs based on both
SDSS Equatorial imaging, as well as new imaging from the 2.6m VLT
Survey Telescope (VST).  AAOmega retains the fibre-fed multi-object
capability across a wide field-of-view from the old 2dF instrument but
the top-end spectrographs have been replaced with a new single bench
mounted spectrograph, with a  red and a blue arm.  \citet{Sharp06}
gives complete instrument details.  In this paper we present the
results from an AAOmega LRG redshift survey. Although the primary
driver for this survey is as a ``Pilot'' study to investigate the
nature of dark energy at high redshift via the BAOs, there are also
several other areas of interest.  By comparing clustering results at
$1 < r < 10 \hmpc$ scales from low ($z<0.4$), intermediate ($z=0.55$),
and high ($z\sim0.7$), redshift LRG studies \citep[][and this study
respectively]{Zehavi05a, Ross07}  we can begin to learn about the
formation and evolution  of the most massive galaxies, and hence,
potentially the most massive dark  matter haloes, from high redshift.

The layout of the paper is as follows.  In Section 2 we describe the
selection criteria used to select our high redshift LRGs.  In Section
3 we give a brief overview of the instrument set-up used and report on
the redshift statistics for our survey, including example spectra.  In
Section 4 we present our clustering results  and in Section 5 we
discuss our results in the context  of other recent results using a
simple Halo Occupation  Distribution (HOD) model.  We conclude in
Section 6.  We assume a flat $\Lambda$CDM cosmology, with  ($\Omm,
\Omlam$)=(0.3,0.7) throughout, unless otherwise explicitly stated.  We
quote distances in terms of~$\hmpc$, where $h$ is the  dimensionless
Hubble constant such that $\ho=100h\kmsmpc$.

\section{SDSS LRG Selection}

At its heart the AAOmega LRG Pilot relies on single-epoch photometric
data from the SDSS \citep{York00, Gunn06} to provide targets for the
recently commissioned AAOmega instrument on the 3.9m Anglo-Australian
Telescope (AAT).

The target selection was designed to select high-redshift LRGs out to
$z\simeq1$ with a mean redshift of $z\simeq0.7$.  Using the SDSS Data
Release 4 \citep[DR4;][]{Adelman-McCarthy06},  we extracted
photometric data for objects classified as galaxies. Three different
selections were then applied to the downloaded data, with the
selections being designed to recover a target sky density of $\sim90$
objects per square degree.

First, we repeat the $gri$-band based selection that was used in the
2SLAQ LRG Survey. We will not repeat the full selection criteria here
(the reader is referred to \citet{Cannon06} for further details)  but
note that LRGs are selected in the $(g-r)$-$(r-i)$ colour-colour plane
with $17.5 < i_{\rm deV} < 19.8$, where $i_{\rm deV}$ is the $i$-band
de Vaucouleurs magnitude.

\begin{figure}
  \includegraphics[height=12.0cm,width=8.0cm]
  {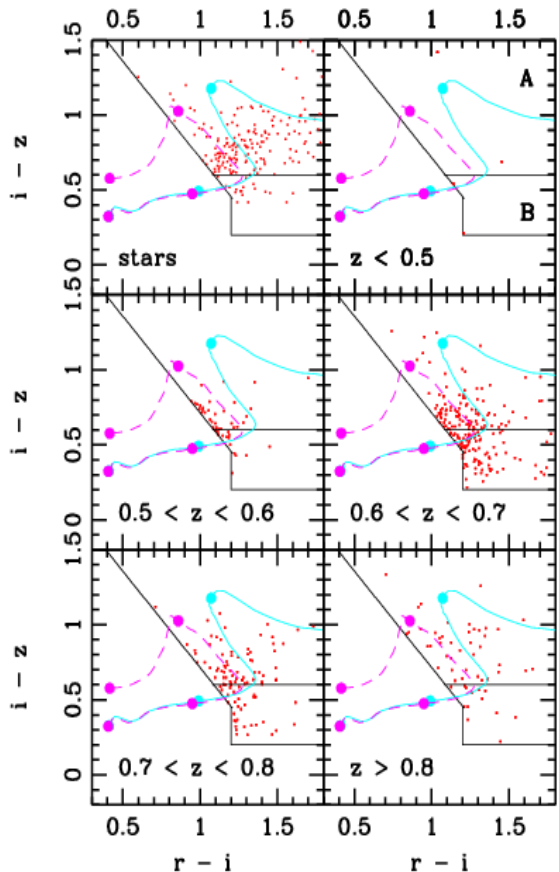} 
  \centering
  \caption[The selection of $z \sim 0.7$ LRGs using the SDSS $riz$-bands]
          {The selection of $z \sim 0.7$ LRGs using the SDSS $riz$-bands. 
	  The (red) dots are objects with confirmed spectroscopic redshifts
	  for both the $19.8 < i_{\rm deV} < 20.5$ and $19.5 < z < 20.2$
	  magnitude selections.	  
	  The tracks are Bruzual \& Charlot models, details given in
	  the text with the solid (cyan) line being a ``single burst'' model 
	  and the dashed (magenta) line having being a $\tau$=1 Gyr model.
	  The diagonal lines are $e_{\parallel}=2.0$.
	  The area labelled ``A'' in the top right redshift $z<0.5$ panel
	  gives the colour-colour space for the higher priority sample, 
	  while area ``B'' is for the lower priority sample.
	  } 
  \label{fig:riz_plane}
\end{figure}

Now with the aim of measuring significantly higher redshifts than the
2SLAQ LRG Survey ($\bar{z}_{\rm 2SLAQ}=0.55$), two further selections
were carried out, this time in the $(r-i)$-$(i-z)$  colour-colour
plane.  The first $riz$-selection had objects in the magnitude range
$19.8 < i_{\rm deV} < 20.5$,  while the second $riz$-selection had
objects  in the magnitude range $19.5 < z < 20.2$, where $z$ is the
SDSS ``Model'' magnitude \citep{Fukugita96, Stoughton02}.  These
magnitude ranges were based on experience gained from the 2SLAQ LRG
Survey as well as the expected performance of the new AAOmega
instrument, such  that LRGs with a significantly higher redshift than
the previous survey could  be selected and observed in a relatively
short exposure ($\sim1.5$ hours).  Within these two $riz$-band
selections, objects were assigned different observational
priorities. The line ``$e_{\parallel}$''was defined (continuing on
from, but not directly related to $c_{\parallel}$ in
\citet{Eisenstein01}  and $d_{\parallel}$ in \citep{Cannon06}), as
\begin{equation}
    e_{\parallel} =  (i-z) +  \frac{9}{7}(r-i)   \ge 2.0.
    \label{eqn:epara}
\end{equation}
and is used to define a boundary in the $riz$-plane.  (All colours
reported here, such as those given in Equation~\ref{eqn:epara}, are
again based on ``Model'' magnitudes).  A higher priority $riz$-plane
cut was imposed with
\begin{equation}
   0.5 \le (r - i) \le 1.8,
\end{equation}
\begin{equation}
   0.6 \le  (i - z) \le 1.5,
\end{equation}
\begin{equation}
   e_{\parallel} \ge 2.0.
\end{equation}
The lower priority cut has
\begin{equation}
   0.2 \le  (i - z) \le 0.6,
\end{equation}
\begin{equation}
   x \le (r - i) \le 1.8,
\end{equation}
where $x$ was the smaller of $e_{\parallel}$ and 1.2 at the given
$(i-z)$.  These cuts can be seen in Figure~\ref{fig:riz_plane} where
the two priorities are shown by the regions marked A and B. The two
evolutionary tracks in Figure~\ref{fig:riz_plane} the stellar
population  synthesis code based on \citet{BC03}. The solid line being
a ``single burst'' model, where  star formation occurs in a single
instantaneous burst at high redshift and then has the stellar
population evolving passively. The dashed line on the other hand is
based on a  model with continuous star formation, with the timescale of
star formation given as $\tau$ = 1 Gyr, where $\tau$ is a decay
constant in that the star formation rate (SFR) is $\propto
\exp^{-t/\tau}$. Both models assume a Salpeter IMF \citep{Salpeter55}
with solar metallicity and a galaxy formation redshift of $z_{\rm form}
= 10$. The evolutionary tracks start near $(r-i) = (i-z) = 0.4$ for
zero redshift, turn upwards near $(r-i) = 1.3$  corresponding to
redshift $z=0.7$ and then turn down again near $(i-z) \sim 1.1$
corresponding to redshift $z =1.0$.  These turning points correspond to
the CaII H+K 4000\AA \, break moving into the $i$- and $z$-bands
respectively.  The solid circles show the colour evolution at redshift
$z=$0.0, 0.5, 1.0 and 1.5.

\begin{table*}
  \begin{center}
    \begin{tabular}{lccrccccrcccrr}
      \hline
      \hline
      Field              & \multicolumn{3}{c}{COSMOS} &       & \multicolumn{4}{c}{COMBO-17 S11}  &        & \multicolumn{3}{c}{2SLAQ d05} & {\bf Survey}\\
      Selection          & $gri$ & $i<20.5$ & all       & \vline & $gri$ & $i<20.5$ & $z<20.2$ & all   & \vline & $gri$ & $i<20.5$ & all  & {\bf  total}\\   
      \hline  
      Spectra Obtained   &  98   &  223  & 321       & \vline & 70    & 262   & 271      & 603   & \vline & 68    & 278   & 346           & {\bf 1270}\\
      $Q{\rm op} \geq 3$ &  71   &  129  & 200       & \vline & 61    & 163   & 143      & 367   & \vline & 57    & 180   & 237           & {\bf  804}\\ 
      LRGs               &  67   &   89  & 156       & \vline & 55    & 119   &  80      & 254   & \vline & 50    & 127   & 177           & {\bf  587}\\
	  \hline
	  \hline
	  \label{tab:Target_Statistics}
    \end{tabular}
    \caption[Redshift Statistics for the AAOmega LRG Pilot Run]
            {Redshift Statistics for the AAOmega LRG Pilot Run.
             These statistics are for the total exposure times 
             as given in Table 1. }
  \end{center}
\end{table*}

\begin{table*}
  \begin{center}
    \begin{tabular}{cccc}
      \hline 
      \hline 
      LRG Sample/ Field (Seeing) & d05 ($1.''6$) & S11 ($1.''8$)  & COSMOS ($2.''1$) \\ 
      \hline 
      $gri$ $i<19.8$ (2SLAQ)     &  $88\pm19$  & $70\pm22$ & $64\pm24$    \\ 
      $riz$ $19.8<i<20.5$       &  $84\pm13$  & $60\pm11$ & $50\pm9$     \\
      \hline 
      \hline
    \end{tabular}
    \caption [LRG percentage  redshift completeness rates]  
	     {LRG percentage  redshift completeness rates
	       ($Q{\rm op} \ge3$) as estimated for $\simeq80$ unfringed
	       fibres between fibres  200-299 in a 1.67hr exposure (stars
	       excluded).   Better observing conditions (d05) yield
	       completenesses  consistent with 2SLAQ. Poorer observing
	       conditions (S11 and  COSMOS) yield lower completeness.
	       The COSMOS data had average airmass 1.4 plus some cloud,
	       as well as poorer seeing.}
      \label{tab:AAOmega_completeness}
  \end{center}
\end{table*}

\section{AAOmega Spectroscopy}

\subsection{Observational Details}

Observations were made on the nights of 03 March 2006 to 07 March 2006
inclusive; the first three nights were Dark nights, the last two were
Grey nights. Of these nights, a total of $\simeq 2$ were lost to
cloud and seeing was frequently poor on the others (see
Table~\ref{tab:The_AAOmega_fields}).  We observed in 3 fields, with
a total area of $\simeq10$ deg$^{2}$, including the COSMOS
field \citep{Scoville07}, the COMBO-17 S11 field \citep{Wolf03} and a
previously observed 2SLAQ Survey field, d05 \citep{Cannon06}, the
coordinates of which are also given in
Table~\ref{tab:The_AAOmega_fields}. For reference, the COSMOS  Survey
has an area of 2 deg$^{2}$, the COMBO-17 S11 field is 0.26
deg$^{2}$ in coverage, while the 2SLAQ LRG Survey has an effective area
of 135 deg$^{2}$ \citep[Sec. 7.2,][]{Cannon06}.

All data were taken with the same spectrograph set-up. The 5700\AA \,
dichroic was used. For the red arm spectrograph the 385R grating was
centred at 7625\AA;  for the blue arm spectrograph the 580V grating
was centred at 4800\AA.  However, no blue arm data was used in our
analysis as the S/N was low, as expected for red galaxies.

Data reduction was performed using the 2dF data reduction pipeline
software, {\tt 2dfdr} \citep{Bailey05} and the redshifts were derived
using {\sc Zcode} developed by Will Sutherland and others for the
2dFGRS Survey \citep[][and references therein]{Colless01}.  The
modifications to {\sc Zcode} originally made for the higher redshift
$z\sim0.5$ galaxies in the 2SLAQ LRG Survey were
retained.  The final catalogue from the AAOmega LRG Pilot contains 1270
unique galaxy spectra with 804 objects having reliable  ``$Q{\rm op}
\geq 3$''\footnote{``$Q{\rm op}$'' represents an integer redshift quality  flag
assigned by visual inspection of the galaxy spectrum  and the redshift
cross-correlation function.  A value of 3 or greater represents a
$>95$\% confidence  that the redshift obtained from the spectrum is
valid.}  redshifts, see Table~\ref{tab:Target_Statistics}.  Of these,
217 objects had M-type stellar spectra leaving 587 high-redshift LRGs.
The COSMOS field contributed 156 LRGs out of 321 obtained spectra,  the
2SLAQ d05 field 177/345 and the S11 field 254/604.  The greater number
of spectra obtained in S11 was due to the fact that objects in the
field were targeted not only with the $19.8 < i < 20.5$  selection but
also with the $19.5 < z <20.2$ $z$-band selection.

We present the catalogue for the first 40 objects in ascending RA in
Appendix A, with the entire catalogue to be published online with the
publication of this paper. In the next Section we report in more detail
on the properties of the high-redshift LRGs.

\subsection{Redshift Completeness}
\begin{figure}
  \includegraphics[height=12.0cm,width=8.5cm]
                  {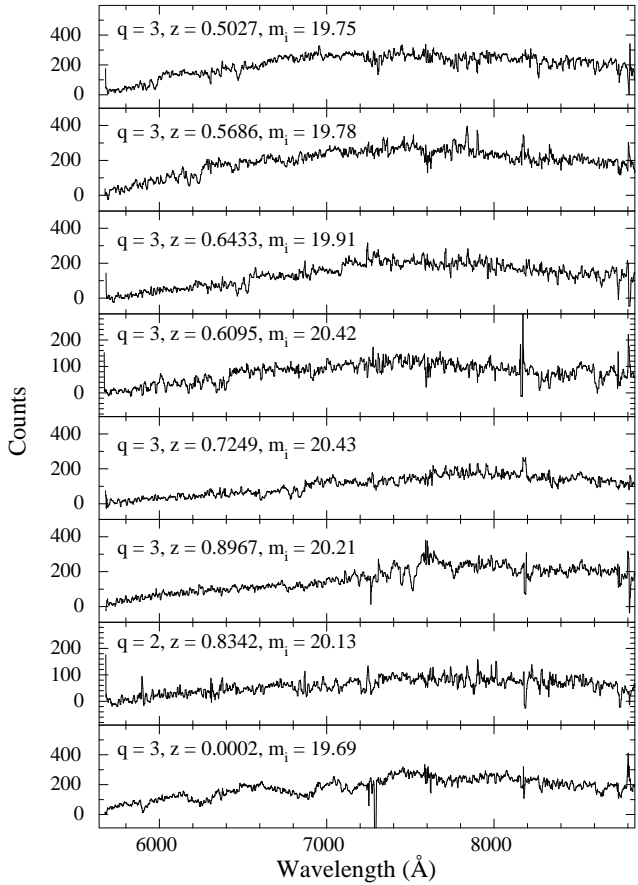}
  \centering
  \caption[Examples of typical AAOmega spectra in 1.67hr exposures, 
           from the $riz$ selected, $19.8 < i < 20.5$ LRG sample.]
          {Examples of typical AAOmega spectra in 1.67hr exposures, 
           from the $riz$ selected, $19.8 < i < 20.5$ LRG sample. 
           The top six panels show spetra of confirmed, $Q{\rm op} \geq 3$ 
           LRGs, with ranging magnitudes and redshifts. 
           The second bottom panel shows an unconfirmed, $Q{\rm op} < 3$, 
           spectrum, while the bottom spectrum is for a confirmed 
           stellar source. }
	  \label{fig:example_spectra_8}
\end{figure}
The LRG redshift completeness statistics for each field can be
calculated from Table~\ref{tab:Target_Statistics} for the full,
$\approx$ 4 hour, exposures and are given in
Table~\ref{tab:AAOmega_completeness}  for a subset of data using 1.67
hour exposures. Our overall completeness was relatively low, 
compared to the 2SLAQ LRG Survey \citep{Cannon06}, but one of the main
reasons for this was due to the several technical issues associated
with the new AAOmega instrument, which have since been corrected.
When checks were made on the d05 field, we found that the redshift
completeness rates for our $riz$, $19.8 < i_{\rm deV} < 20.5$ targets
as estimated from $\approx 80$ ``unfringed'' fibres were $90\pm9\%$ in
$\approx$4 hour exposures, $84\pm13\%$ in 1.67 hour exposures in
1.$''$6 seeing.  Thus, using the full number of sub-exposures we found
no significant increase in redshift completeness compared to a 1.67
hour exposure, although this may still be due to conditions varying
within the 3 hour exposure time.  But our general conclusion is that
with reasonable seeing and transparency, we achieve 85-90\% redshift
completeness in a 1.67 hour exposure. We show a selection of spectra
from the subset of data taken in the d05 field in
Figure~\ref{fig:example_spectra_8}. The top six panels show spetra of
confirmed, $Q{\rm op} \geq 3$ LRGs, with ranging magnitudes and
redshifts, including a high redshift confirmed LRG at
$z\approx0.9$. The second bottom panel shows an unconfirmed, $Q{\rm
op} < 3$, spectrum, while the bottom spectrum is for a confirmed
M-star. The improved AAOmega throughput and sky subtraction enables us
to work further into the near-infrared, allowing us to  probe higher
redshifts. Note the prominent CaII H+K 4000\AA \, break appears in all 
the confirmed spectra, as expected for an old stellar population. 

We also confirmed that the exposure time needed to obtain reliable
redshifts of LRGs selected in the same manner as the 2SLAQ survey
(using a $gri$-band, $i<19.8$ selection) was cut by a factor of $\sim
4$ from the old 2dF instrument. We note from Table 3 that at least in
the more reasonable observing conditions for  the d05 field that the
completeness of the 1.67hr LRG sample is consistent with the high,
90\%, completeness achieved for 2SLAQ LRGs.

\subsection{Redshift Distribution}

\begin{figure}
  \includegraphics[height=6.5cm,width=6.5cm]
                  {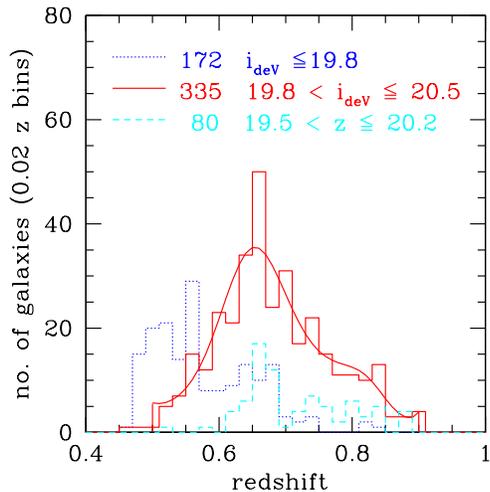}
  \centering
  \caption[The $N(z)$ of the AAOmega LRG Pilot]
          {The $N(z)$ of $Q{\rm op} \geq 3$ LRGs from the AAOmega LRG Pilot Run,
            showing that $0.5\leq z \leq0.9$ can be readily selected using 
            SDSS $riz$ photometry.
	     The dotted (blue) histogram shows the distribution for the 
             $i_{\rm deV} < 19.8$ $gri$-selection, 
             while the solid (red) and the dashed
	     (cyan) histograms show the $riz$ selections with  
	     $19.8 < i_{\rm deV} < 20.5$  and $19.5 < z < 20.2$ respectively.
             We also plot the polynomial fit (red line) that is used to
             model the $N(z)$ distribution for the $riz$, 
             $19.8 < i_{\rm deV} < 20.5$ selection in Section 4.2.} 
	  \label{fig:AAOmega_Nofz}
\end{figure}
\begin{figure}
  \includegraphics[height=6.5cm,width=6.5cm]
                  {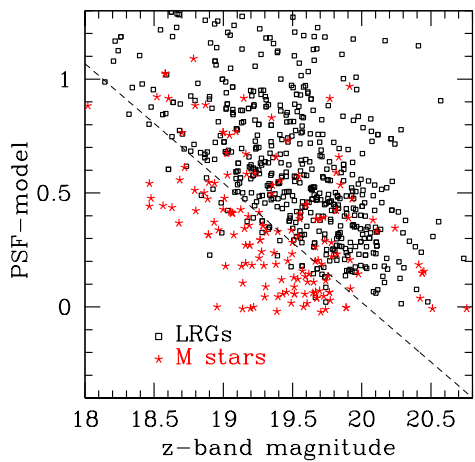}
  \centering
  \caption[Star-Galaxy Separation using SDSS $z$-band magnitudes]
          {Star-Galaxy Separation using SDSS $z$-band magnitudes.
	    All objects with $Q{\rm op} \geq 3$ and $19.8 < i_{\rm deV} <20.5$ 
	    are shown, with objects having stellar spectra plotted as (red) 
	    stars and objects having high-redshift LRG spectra plotted 
	    as (black) open  squares. 
           The ordinate gives the difference between the ``PSF'' and 
          ``Model'' $z$-band magnitudes as given from the SDSS DR4 imaging.}
	  \label{fig:star_gal}
\end{figure}

The {\it raison d'$\hat{e}$tre} of the AAOmega LRG Pilot run was to
test if we could readily select $z\sim 0.7$ LRGs using single-epoch
SDSS $riz$-photometry.  As can be seen in
Figure~\ref{fig:AAOmega_Nofz},  where we plot the redshift
distributions for confirmed $Q{\rm op} \geq 3$ LRGs,  this proved
feasible.  The mean redshift of our $19.8 < i_{\rm deV} < 20.5$
magnitude sample was  $z=0.681\pm 0.005$, with a strong tail out to
redshift $z=0.8$ and indeed some  objects at $z=0.9$. We found that
there was no major difference between the samples with different
priorities (areas ``A'' and ``B'' in Figure~\ref{fig:riz_plane}).
Also shown in Figure~\ref{fig:riz_plane} are the $riz$-band colours
for the objects with spectroscopically confirmed redshifts.  When the
magnitude limits applied were changed from $19.8 < i_{\rm deV} < 20.5$
to $19.5 < z < 20.2$, the mean redshift increased to $z = 0.698 \pm
0.015$. The mean redshift for our $gri$-band, $17.7 < i_{\rm deV} <
19.8$ selection was very comparable to the 2SLAQ LRG Survey at
$z=0.578\pm0.006$.

However, since we found that even though we were able to obtain LRG
spectra for $z<20.2$ objects from SDSS single-epoch imaging (and get the
increase in redshift one might expect based on galaxy colours from
evolutionary models), we find that the completeness of this
sample dropped significantly and longer, $\geq2$ hours,
exposures would  be required in order to obtain $Q{\rm op} \geq 3$
redshifts. This is not surprising considering that with a $z < 20.2$
magnitude limit, we are selecting objects with $i_{\rm deV}\sim$20.8
given a $(i-z)$ colour of $\sim$0.6 (as seen in Fig. 1). Thus for the
remainder of this analysis, and the eventual strategy for a large
LRG-BAO Survey, we only consider objects with $19.8 < i_{\rm deV} <
20.5$.  

As can be seen from Table~\ref{tab:Target_Statistics}, a significant
fraction ($27\%$)  of our $Q{\rm op} \geq 3$ objects were M-type
stars. However, as shown in Figure~\ref{fig:star_gal}, {\it a
posteriori} checking  shows that we can reject 40\% of these stars
using a star-galaxy separation in the  $z$-band, rather than the
standard SDSS separation performed in the $r$-band. The stellar
contamination drops to $16\%$, with very few high-redshift galaxies
being lost. Employing near-IR imaging data, specifically a $J-K >1.3$
cut, would dramatically reduce the stellar contamination further, to
the levels of a few percent.

\subsection{2SLAQ, COMBO-17 and AAOmega Comparison}
\begin{figure}
  \includegraphics[height=6.5cm,width=6.5cm]
		  {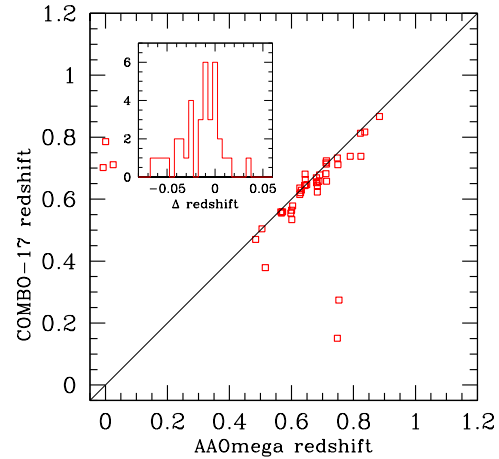}
  \centering
  \caption[COMBO-17 photometric redshifts vs. 
           AAOmega spectroscopic redshifts.]
          {COMBO-17 photometric redshifts vs. 
	    AAOmega  spectroscopic redshifts. 
           The solid line is the 1:1 relation. 
	   The insert shows the histogram of
	   $\Delta z = z_{\rm spec} - z_{\rm phot}$
	   for AAOmega and COMBO-17 redshifts
	   respectively.}
  \label{fig:COMBO17_comparison}
\end{figure}

In Figure~\ref{fig:COMBO17_comparison} we show a comparison between the
spectroscopic redshifts we recorded from our AAOmega observations and
those measured photometrically by the Classifying Objects by Medium-Band
Observations (COMBO-17) survey \citep[e.g.][]{Wolf03, Bell04a,
Phleps06}. As can be seen, the 43 common photometric and spectroscopic
redshifts match extremely well for the objects for which we have secure
redshifts ($Q{\rm op} \geq 3$). There seems to be a slight trend for
the photometric redshifts to underestimate the spectroscopic
redshift. Why this is the case is not well understood. Excluding 5
``catastrophic failures'', where $| \Delta z | \geq 0.2$, the average
offset between the COMBO-17 photometric and AAOmega spectroscopic
redshifts is $\overline{\Delta z}=0.026  \pm 0.005$, in the sense that
COMBO-17 redshifts are too small.  There are 3 spectroscopically
confirmed stars that COMBO-17 classified as redshift $z\sim 0.7$ galaxies.

We also compare the spectroscopic redshifts measured by AAOmega with
those obtained in the 2SLAQ LRG Survey.  We find, for the $Q{\rm op} \geq 3$
LRGs common in both, the mean $\Delta z = 8.4\times10^{-4}$ with the
spread on the difference in redshifts being $1.24\times10^{-3}$  i.e. $
370 \kms$. If the error is split evenly between the two surveys, then
the error on AAOmega LRG redshifts is $\pm \, 370/\sqrt{2} = \pm 260 \kms$.


\section{LRG Clustering Results}
\subsection{AAOmega LRG Angular Correlation Function, $w(\theta)$}
\label{sec:AAOmega_wtheta}
\begin{figure}
  \includegraphics[height=6.5cm,width=6.5cm]
		  {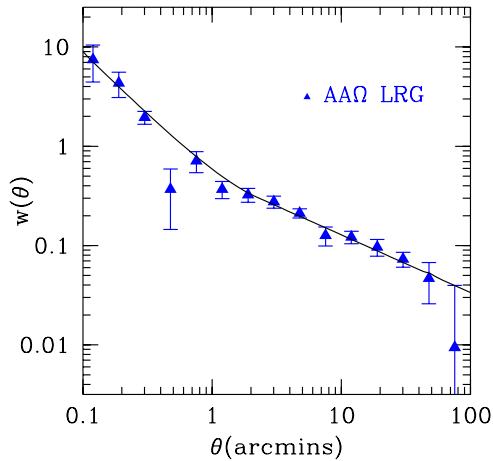}
		  \centering
  \caption[The AAOmega LRG Pilot angular correlation function,
          $w(\theta)$] 
          {The AAOmega LRG Pilot angular correlation function,
          $w(\theta)$,  is given by the solid (blue) triangles.
           \hbox {2 326} objects were used with magnitudes in the 
           range $19.8 < i_{\rm deV} < 20.5$.
          The solid (black) line is a estimation of $w(\theta)$  given our
          redshift distribution and projecting using Limber's
          Formula, with the associated $r_{0}$ and $\gamma$ 
          jackknifed values given in Table~\ref{tab:w_theta_values}.}
  \label{fig:AAOmega_w_theta}
\end{figure}
Using the procedure described by \citet{Ross07}, the projected angular
correlation function, $w(\theta)$, for the AAOmega LRG Pilot Survey is
presented in Figure~\ref{fig:AAOmega_w_theta}. The solid (blue) triangles
are for the measurements made utilising the  ``Input Catalogue'' from
which objects were selected as potential high-redshift LRG
candidates. Approximately \hbox{2 300} objects were used in this
measurement from 6 fields that were observed by the 2SLAQ Survey,
each $\pi$ deg$^{2}$ in area. All these objects were 
potential targets having passed the $riz$-cuts discussed above. 
Field centres of the 6 fields are given in Table~\ref{tab:w_theta_fields}.
It should also be noted that the star-galaxy separation discussed 
above was applied to this input sample. 
\begin{table}
  \begin{center}
    \begin{tabular}{rccc}
      \hline
      \hline
      Field Name & R.A. (J2000) & DEC (J2000) \\ 
      \hline
      2SLAQ  c05  & 12h 38m 18s  & -00 12 35 \\ 
      \, " \,\,\, c07  & 12h 47m 54s  & -00 12 35 \\ 
      \, " \,\,\, d07  & 13h 31m 12s  & -00 12 35 \\ 
      \, " \,\,\, e01  & 14h 34m 00s  & -00 12 35 \\ 
      \, " \,\,\, e03  & 14h 42m 48s  & -00 12 35 \\ 
      \, " \,\,\, c07  & 12h 47m 54s  & -00 12 35 \\ 
      \hline
      \hline
    \end{tabular}
  \end{center}
  \caption[Details of the 2dF fields that were used for the $w(\theta)$ 
  measurements. Note, d05 was also used and details of this field 
  are given in Table 1.]
  {Details of the 2dF fields that were used for the $w(\theta)$ measurements.
    Note, d05 was also used and details of this field are given in Table 1.
    All 6 fields were observed by the 2SLAQ Survey. }
  \label{tab:w_theta_fields}
\end{table}
The error bars associated with the AAOmega LRG $w(\theta)$ measurement
are {\it field-to-field} errors \citep[see][]{Ross07} and do not  take
into account the fact that the clustering measurements  are correlated
and therefore, the errors on these points  should only be regarded as
indicative. When we come to calculate the errors on the  fitted
power-law parameters, defined in equation~\ref{eqn:pl_fit},  we
perform a jackknife analysis on our measurements in the attempt to
take into account these covariances.  This involves removing one field
at a time from our sample and   recomputing and refitting the angular
correlation  function, weighting by the number of $DR$ pairs.  As
such, we present these jackknife errors  for our measurements in
Table~\ref{tab:w_theta_values}. 
\begin{table}
  \begin{center}
    \begin{tabular}{lrrrr}
      \hline
      \hline
                       &   2SLAQ LRG        & AAOmega LRG \\
\hline
$r_{0, \rm ss} /\hmpc$  &  5.47$\pm$0.40    &  5.0$\pm$0.34      \\
$\gamma_{\rm ss}$       &  2.16$\pm$0.07    & 2.28$\pm$0.04     \\
\hline
$r_{0, \rm ls} /\hmpc$ &  8.0$\pm$0.8\phantom{0}    &   10.2$\pm$0.7\phantom{0}    \\
$\gamma_{\rm ls}$      &  1.67$\pm$0.07   &   1.58$\pm$0.09    \\
\hline
\hline
      \end{tabular}
\end{center}
\caption[The values of $r_{0}$ and $\gamma$ for the 2SLAQ LRG Survey and
AAOmega LRGs.]{The values of $r_{0}$ and $\gamma$ for the 2SLAQ LRG Survey and
AAOmega LRGs. Note that $r_{b}=1.5 \hmpc$ for the 2SLAQ LRGs, while $r_{b}=1.0 
\hmpc$ for AAOmega LRGs. Also note that due to improved implementation of
Limber's formula and more accurate binning, the values given here 
for $r_{0}$ and $\gamma$ for the 2SLAQ LRG Survey from Limber's Formula, 
supersede those given by \citet{Ross07}.} 
\label{tab:w_theta_values}
\end{table} 

A single power-law, of the form
\begin{equation}
  \xi(r) = \left ( \frac{r}{r_{0}} \right)^{-\gamma},
 \label{eqn:pl_fit}
\end{equation}
where $r_{0}$ is the correlation length and $\gamma$ the power-law slope, 
has traditionally been fitted for the 3-D correlation function for galaxies, 
$\xi$, and from which the relation,  
\begin{equation}
  w(\theta) = A \, \theta^{1-\gamma}
 \label{eqn:pl_fit_w_theta}
\end{equation}
where $A$ is amplitude, can be derived for the angular
correlation function (e.g. Peebles, 1980).  
However, as was also found by \citet{Ross07} for the 2SLAQ LRG
$w(\theta)$,  here we find that a double power-law model is required to fit
the present  measurement. Following that work, we use Limber's Formula
\citep[see][]{Phillipps78} to relate the 3-D correlation function to
the our measured $w(\theta)$.   A double power-law of the form
\begin{equation}
  \xi(r)= \left\{ 
   \begin{array}{ll}
     \left(r / r_{0, \rm ss}\right)^{-\gamma_{\rm ss}} & r \leqslant r_{\rm{b}}
                                            \;\;\;  \rm{and}\;\; \\
     \left(r / r_{0, \rm ls}\right)^{-\gamma_{\rm ls}} & r  > r_{\rm{b}}
   \end{array}
        \right.
  \label{eq:xipowerlaw2}
\end{equation}
where `ss' and `ls' stand for small scales and large scales
respectively, is assumed and calculated from Limber's formula. 
The calculated values for $r_{0}$ and $\gamma$ 
are given in Table~\ref{tab:w_theta_values}, 
where we fit over the range $0.1' < \theta < 40.0'$
and note that $r_{b}=1.5 \hmpc$ for the 2SLAQ LRGs, while $r_{b}=1.0 
\hmpc$ for AAOmega LRGs. We also note that due to improved implementation of
Limber's formula and more accurate binning, the values given here for $r_{0}$ 
and $\gamma$ for the 2SLAQ LRG Survey from Limber's Formula, supersede those
given by \citet{Ross07}.  

From Table~\ref{tab:w_theta_values}, we can see that the $w(\theta)$
measurement for the AAOmega high-redshift data is comparable to the
$z=0.55$ data from the 2SLAQ LRG survey. At small scales, the
observed AAOmega $w(\theta)$ slope is  nearly equal to the 2SLAQ LRG
measurement,  while at large-scales, the AAOmega slope is slightly
shallower than the 2SLAQ LRGs: $\gamma=1.58\pm0.09$ for AAOmega
compared to $\gamma=1.67\pm0.07$  for 2SLAQ.  However, given the
associated errors, the two measurements are in very good agreement. We
leave further analysis of the angular correlation function as reported
here to Sawangwit et al. (2008, in prep.) who shall investigate the
evidence for a double power-law feature in a much larger LRG sample. 

Given the AAOmega LRG Pilot $N(z)$ (Figure~\ref{fig:AAOmega_Nofz}) and
using Limber's Formula, the AAOmega $w(\theta)$ amplitude is expected
to be 13\% lower than the 2SLAQ LRG amplitude if there is no
clustering evolution in comoving coordinates. Thus, in terms of the
overall amplitude, this reinforces the impression given in Table 5
that AAOmega LRGs have a large-scale amplitude which is at least as
high as the 2SLAQ LRGs.  This finding is further backed up by
measurements of the projected correlation function,
$w_{p}(\sigma)$. We do not present our $w_{p}(\sigma)$ results here,
but note that our best fitting (single) power-law to this data  has an
amplitude $r_{0}=9.0\pm0.9 \hmpc$ and slope $\gamma=1.73\pm0.08$ over the
scales $1.0 < \sigma/ \hmpc < 40.0$ (where $\sigma$ is the separation
across the line-of-sight).

\subsection{Redshift-space Correlation Function, $\xi(s)$}

Using the spectroscopic redshift data we obtained in the COSMOS, S11 and
d05 fields we now calculate the 3-D redshift-space correlation function, $\xis$.
We use the minimum variance estimator suggested by \citet{LS93}
(proven to be an optimum estimator by \citet{Kerscher00}) where  
\begin{eqnarray}
  \xi(s) &=& 1+\left( \frac{N_{rd}}{N} \right)^{2}
             \frac{DD(s)}{RR(s)}
             -  2   \left( \frac{N_{rd}}{N} \right)
             \frac{DR(s)}{RR(s)}
\label{lseq}
\end{eqnarray}
and $DD$, $DR$ and $RR$ are the number of data-data, data-random and
random-random pairs at separation $s$ respectively. We use bin widths of
$\delta\log(s / \hmpc)$=0.2 and the number density  of random points
was 20$\times$ that of the LRGs.

The random catalogue was made taking into account the angular
incompleteness and the radial distribution of the objects in this
Pilot. For each 2dF field we constructed a ``quadrant bullseye'' angular
mask which consisted of 5 concentric rings divided into 4 quadrants.
Using both the input catalogue and the 2dF instrument
configuration positions, a completeness map was made in each of the 20
sectors. These completenesses then went into mimicking the angular
selection function, from which a random catalogue was generated. 
Corrections for fibre collisions on small, $\lesssim 30$ arcseconds, 
scales were made by taking the ratio of the input catalogue $w(\theta)$ 
to the observed redshift catalogue $w(\theta)$, as described by \citet{Ross07}.  
The radial distribution was described by a high-order polynomial fit
(shown as the red curve in Figure 3) to the
AAOmega $N(z)$ for the 335 $19.8 < i <20.5$ selected LRGs given in
Figure 3. We also note that for ease of modelling, 
we truncate the polynomial fit (and thus the random radial distribution)
at redshifts of $z \leq 0.50$ and $z \geq 0.90$.
\begin{figure}
    \includegraphics[height=6.5cm,width=6.5cm]
                    {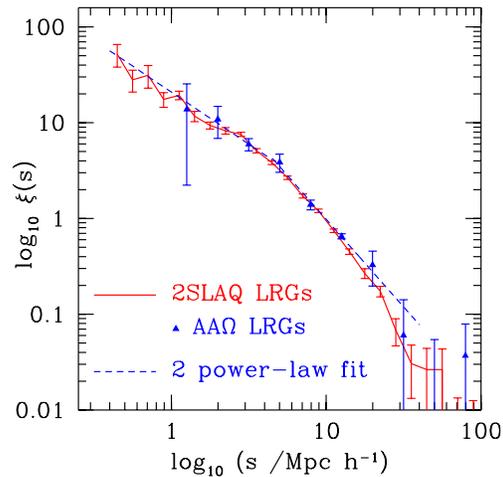} 
   \centering
   \caption[The AAOmega LRG Pilot Redshift-Space Correlation Function $\xi(s)$]
	   {The AAOmega LRG Pilot Redshift-Space Correlation Function $\xi(s)$.
	     The (blue) triangles are the measurements from 
             the $riz$-selected $19.8 < i_{\rm deV} < 20.5$ sample, 
             which yielded 335 $Q{\rm op} \geq 3$ LRGs and
             the associated ``Field-to-Field''  errors. 
             The dashed (red) line is the redshift-space correlation 
	     function from the 2SLAQ LRG Survey \citep{Ross07}.}
  \label{fig:AAOmega_xis}
\end{figure}

Figure~\ref{fig:AAOmega_xis} shows our estimate of the 3-D
redshift-space  correlation function, $\xi(s)$. Again, our error
estimates are based on ``field-to-field'' errors.  For $\xi(s)$, we
use a double power-law model of the form given in
equation~\ref{eq:xipowerlaw2}, motivated by the fact that we expect 
the small-scale correlation function to be smoothed bt the effect of 
velocity dispersion (or ``Fingers-of-God'') whereas at larger scales
we expect the correlation function simply to be boosted due to infall, 
characterised by the parameter $\beta=\Omega^{0.6}/b$.
We adopt the same procedure as for
$w(\theta)$ and do a jackknife error analysis in order to estimate the
errorbars on the best-fit double power-law model parameters. We find
that,  $s_{0, \rm ss}= 16.5\pm4.0 \hmpc$ with $\gamma_{\rm ss}=1.09\pm0.28$
on scales $s < 4.5 \hmpc$ and $s_{0, \rm ls}=  9.9\pm0.7 \hmpc$ with
$\gamma_{\rm ls}=1.83\pm0.35$   on scales $s > 4.5 \hmpc$. The
clustering strength for the $19.8 < i <20.5$, $riz$-selected AAOmega
LRGs is again very comparable to the 2SLAQ LRG Survey, where $s_{\rm
ss}=17.3^{+2.5}_{-2.0} \hmpc$ and $\gamma_{\rm ss}=1.03\pm0.07$  on scales
$s < 4.5 \hmpc$ and $s_{\rm ls}= 9.40\pm0.19 \hmpc$ and $\gamma_{\rm
ls}=2.02 \pm 0.07$ on scales $s > 4.5 \hmpc$.

\begin{table*}
  \begin{center}
    \setlength{\tabcolsep}{4pt}
    \begin{tabular}{lccclcl}
    \hline
    \hline
    Survey         & mean redshift  & $n / h^{3} {\rm Mpc^{-3}}$ & Luminosity &  $\hmpc$  & $\gamma$  & Reference   \\
    \hline   
AAOmega $riz$ LRG  & 0.68      & $\sim2\times10^{-4}$   & $\gtrsim 2L^{*}$ & $r_{0}=$10.2$\pm$0.7   & 1.58$\pm$0.09 & 1 \\
                   &           &                       &                  & $r_{0}=$9.0$\pm$0.9  & 1.73$\pm$0.08 & 2  \\
                   &           &                       &                  & $s_{0}=$9.9$\pm$0.7    & 1.83$\pm$0.35 & 3 \\
2SLAQ LRG          & 0.55      & $\sim2\times10^{-4}$   & $\gtrsim 2L^{*}$ & $s_{0}=$9.40$\pm$0.19  & 1.98$\pm$0.07 & 4, 5 \\
                   &           &                       &                  & $r_{0}=$7.45$\pm$0.35  & 1.72$\pm$0.06 & 4, 5  \\
SDSS LRG           & 0.28      & $9.7\times10^{-5}$     & $\geq 3L^{*}$    & $s_{0}=$11.85$\pm$0.23 & 1.91$\pm$0.07 & 6 \\
                   &           &                       &                  & $r_{0}=$9.80$\pm$0.20  & 1.94$\pm$0.02 & 6 \\
MegaZ-LRG          & 0.63      & $5.6\times10^{-5}$     & $\gtrsim 3L^{*}$ & $r_{0}=$9.3$\pm$0.3    & 1.94$\pm$0.02 & 7 \\
COMBO-17           & 0.6       & $4\times10^{-3}$       & $\sim L^{*}$     & $r_{0}=$5.39$^{+0.30}_{-0.28}$ & 1.94$\pm$0.03 & 8 \\
NDWFS              & $\sim$0.7 & $\approx1\times10^{-3}$ & $>1.6L^{*}$     & $r_{0}=$6.4$\pm$1.5    & 2.09$\pm$0.02 & 9, 10 \\
  \hline
  \hline
    \end{tabular}
    \caption[Values of $s_{0}$ and $r_{0}$ from the \ATLAS LRG Pilot, 
      SDSS LRG Survey and the 2SLAQ LRG Survey.]
            {Values of $s_{0}$ and $r_{0}$ from the \ATLAS LRG Pilot
	      using the $w(\theta)$ measurement, the fit to
	      $w_{p}(\sigma)$  and the $\xi(s)$ calculation with $s >
	      4.5 \hmpc$.  Values from the SDSS LRG Survey 
              ($-23.2 < M_{g} < -21.2$), the 2SLAQ LRG Survey, 
              MegaZ-LRG and the NDWFS are also given.  Note that due
	      to redshift-space distortions and other non-linear
	      effects, $r_{0}$ will usually be smaller than $s_{0}$.
              (1) this work, from $w(\theta)$; (2) this work, from $\wp$; 
              (3) this work, from $\xi(s)$; 
              (4) Ross et al. (2007); (5) Wake et al. (2006);
              (6) \citet{Zehavi05a}; 
              (7) Blake et al. (2007); 
              (8) Phleps et al. (2006); 
              (9) White et al. (2007); (10) Brown et al. (2008).}
  \end{center}
  \label{tab:r_nought_values}
\end{table*}

Using the model of \citet{Kaiser87}, we can find the parameter $\beta$ via
\begin{equation}
\xi(s)= \xi(r) \left({ 1+\frac{2}{3}\beta + \frac{1}{5}\beta^{2}} \right).
\end{equation}
We use our power-law fit for $\xi(r)$ and our large-scale  power-law fit
to $\xi(s)$ and find that the ratio $\xi(s) / \xi(r) = 1.3\pm0.3$
corresponding to  a value of  $\beta \simeq 0.4 $ at a scale of $8
\hmpc$. This is not inconsistent with the value $\beta = 0.45\pm0.05$
found for the 2SLAQ LRGs, though clearly the errorbar is
large. Nevertheless, for a reasonable value of $\beta$, our values of
$s_{0}= 9.9 \pm 0.7\hmpc$ and $r_{0}=9.0\pm0.9 \hmpc$ appear
consistent. These high clustering amplitudes clearly suggest that 
at $z\simeq0.7$, LRGs remain very strongly clustered.

\section{Discussion}

\subsection{ Clustering amplitudes and bias of LRGs at $z\simeq 0.7$}

Now that we have calculated the AAOmega LRG angular,  projected, and
3-D redshift-space correlation functions  we can use these
measurements to infer the physical properties of LRGs.  Before
proceeding to determine typical LRG halo masses using simple `halo
occupation' models, we first compare the clustering amplitudes and
biases of the AAOmega LRGs with other LRG results, taking into account
the different redshift and luminosity ranges.  For reference, a
summary of results of space densities, luminosity limits and
clustering amplitudes from the AAOmega LRG, 2SLAQ LRG, SDSS LRG,
MegaZ-LRG, COMBO-17 and NDWFS surveys, is given in Table 6.  We note,
however, that direct comparisons between clustering results  from
surveys with different e.g. magnitude and colour selections  can be
complex. 

We have found that a 2-power law fit is consistent with AAOmega
$w(\theta)$ data. The slopes of the AAOmega power-law fits are both
less than those for the 2SLAQ LRG Survey \citep{Ross07}.  This could
be due to evolution with redshift but the errors on the AAOmega
$w(\theta)$ are too large for this difference to be significant.
Certainly the large scale results from $\xi(s)$ are perfectly
consistent with the two surveys having the same large-scale slope and
amplitude  (see Fig.~\ref{fig:AAOmega_xis}).

We further note that from both the fitting of Limber's formula to
$w(\theta)$ and describing $w_{p}(\sigma)$ with a simple power-law,
we find the real-space clustering amplitude of AAOmega LRGs is
consistent with that from the SDSS LRG Survey \citep{Zehavi05a},
though our errors are large.  Using our $r_{0}$ estimate from
$w_{p}(\sigma)$, (which has the smaller error and  more closely
matched power-law slope), we note that AAOmega LRGs have a slightly
lower clustering amplitude than SDSS LRGs, $r_{0}=9.0\pm0.9 \hmpc$
versus $r_{0}=9.80\pm0.20 \hmpc$ respectively. However, this is not
surprising since SDSS LRGs have a redder colour selection and higher
luminosity, and this may explain their higher clustering amplitude. 

To calculate the value of the linear bias, $b$, for the AAOmega LRGs, 
we use the integrated correlation function \citep{Croom05, daAngela08}, 
\begin{equation}
\xi_{20}(r) = \frac{3}{r_{\rm max}^{3}} \int^{r_{\rm max}}_{0} \xi(r) r^{2} dr
\end{equation}
where we set $r_{\rm max}=20 \hmpc$ since this is a large enough scale 
for linear theory to apply and also, due to the
$r^{2}$ weighting, small-scale redshift-space distortions should be
negligible.  We first calculate the integrated mass correlation
function using the $\sigma_{8}=0.84$ normalised $\Lambda$CDM model
for $P(k)$ from \citet{Smith03} with $\Omm(z=0)=0.27$.  We find
$\xi^{\rm mass}_{20}=0.12$ at the 2SLAQ LRG mean redshift $z=0.55$
and $\xi^{\rm mass}_{20}=0.11$ at the AAOmega LRG mean redshift
$z\simeq0.70$. 

We then calculate the integrated galaxy correlation function assuming
$r_{0}=7.45\pm0.35 \hmpc$ and hold $\gamma$ fixed at 1.72 for the
2SLAQ LRGs \citet{Ross07} and $r_{0}=9.03\pm0.93 \hmpc$, $\gamma=1.73$
for AAOmega LRGs.  We find that $b_{\rm 2SLAQ} = 1.90\pm0.08$ and
$b_{\rm AAOmega } = 2.35\pm0.22$, where $b=(\xi_{20} / \xi_{\rm mass,
20})^{1/2}$.  The value of $b_{\rm 2SLAQ} = 1.90\pm0.08$ is higher,
but consistent with that found by \citet{Ross07}, who found $b_{\rm
2SLAQ} = 1.66\pm0.35$,  from $z$-space distortion analysis,  and we
suggest the error presented here may be an underestimate since
$\gamma$ is being held at a fixed value.  The value of $b_{\rm AAOmega
} = 2.35\pm0.22$ is higher than for the 2SLAQ LRGs, but the large
error on the AAOmega result means there may be no inconsistency here.
However, our value of $b_{\rm AAOmega } = 2.35\pm0.22$ is even higher
than that reported for the SDSS LRGs at lower redshifts, who report
values of $b\approx1.8$ \citep{Padmanabhan07}.  Although an increase
in bias is expected due to the higher redshift of the  AAOmega sample,
the effect is larger than predicted especially taking into account the
bluer AAOmega selection. But again the large error on $b_{\rm
AAOmega}$ renders this difference statistically insignificant. 

To see what sort of consistency with 2SLAQ might be expected,
we can predict the value of $b$ at redshift $z=0.7$
by utilising the values measured by 2SLAQ at lower redshift,
$b(z=0.55)=1.66\pm0.35$, and  the bias evolution model given
by \citet{Fry96, Croom96},
\begin{equation}
b(z) = 1 + [b(0) - 1] G(\Omm(0), \Omlam(0), z).
\label{eqn:bias_model_1}
\end{equation}

Here, $G(\Omm(0), \Omlam(0), z)$ is the linear growth rate of the
density perturbations \citep{Peebles80, Peebles84, Carroll92}.  There
are many other bias models, but here we are following \citet[][and
references therein]{Ross07} by making the simple assumptions that
galaxies formed at early times  and their subsequent clustering is
governed purely by their discrete motion within the gravitational
potential produced by the matter density perturbations.  This model
would be appropriate, for example, in a  ``high-peaks'' biasing
scenario where early-type galaxies formed at a single  redshift and
their co-moving space density then remained constant to the  present
day. 

Thus, assuming a growth rate of $G(0.3,0.7,z)$,  to relate $\xi_{\rm
mm}(z=0.55)$ to $\xi_{\rm mm}(z=0.7)$,  we therefore expect $\xi_{\rm
gg}(z=0.7) = 0.94 \, \xi_{\rm gg}(z=0.55)$  from this model.  From
Table 6 the $r_{0}$  values between 2SLAQ and AAOmega LRGs are
consistent, although the errors on the AAOmega $r_{0}$ measurement are
big. But the errors on $\xi(s)$ are smaller, and even here, the
$s_{0}$ values agree to  within the errors (see also
Figure~\ref{fig:AAOmega_xis}). The consistency of the clustering
results is expected, since the 0.7 magnitudes deeper $19.8 < i_{deV} <
20.5$ selection was based on experience from the 2SLAQ LRG Survey and
primarily designed  to select similarly highly-biased red galaxies at
redshift $z\simeq 0.7$. We conclude that the LRG correlation function
amplitudes are similar at redshifts $z\approx0.55$ and $z \approx 0.7$
and that there is still no inconsistency with the simple bias model
where the comoving density of LRGs are assumed to be constant with
redshift.

\subsection{Predictions of halo occupation models}

An alternative approach to interpreting our measured level of
clustering is to use the halo occupation model, in which the galaxy
field is taken to be a superposition of contributions from dark-matter
haloes, weighted by the number of galaxies per halo, $N(M)$. 
This methodology is commonly reffered to as a `halo occupation distribution', 
or HOD, model and was used recently by \cite{Phleps06} to model the
projected correlations in the COMBO-17 survey. We apply exactly
the same method as described in that paper to model our AAOmega data,
specifically for our $w_{p}(\sigma)$ measurement.  Again we adopt a 
standard matter power spectrum, with $\Omega_m=0.3$, $\Omega_b=0.045$,
$h=0.73$, $\sigma_8=0.85$,  and a scalar spectral index of 0.97. The
occupation model is the simplest possible: $N(M) = (M/M_{\rm
min})^\alpha$ for $M>M_{\rm min}$.  These two free parameters are
reduced to one if the model is also required to match the number
density of LRGs, which is approximately $0.0002 \, h^3 \, {\rm
Mpc}^{-3}$.

Realistic occupation models will be more complicated than this simple
power-law form, but Phleps et al. argue that the results can be
expressed quite robustly in terms of an effective halo mass --
i.e. the average halo mass weighted by the number of galaxies. For our
current data, the occupation parameters that best match the clustering
measurements are $\alpha\simeq 0.7$ and $M_{\rm min}\simeq 2 \times
10^{13} h^{-1}M_\odot$.  These imply an average halo mass for the
AAOmega LRGs at $z\simeq 0.7$ of $M_{\rm eff}\simeq 7\times 10^{13}
h^{-1} M_\odot$.  Reasonably enough for particularly rare and luminous
galaxies such as those studied here, this mass is somewhat larger than
the figure found by Phleps et al. for the COMBO-17 red-sequence
galaxies at $z\simeq 0.6$, which was $M_{\rm eff}\simeq 1.6\times
10^{13} h^{-1}M_\odot$, using the same methodology. Our AAOmega figure
for $M_{\rm eff}$ is in fact almost identical to the average mass
deduced for $z=0$ red-sequence galaxies in SDSS. Of course, this
coincidence does not imply any direct correspondence between these
populations: the haloes that host our $z\simeq0.7$ LRGs may have
become much more massive by the present. 

\citet{Blake07} calculate the LRG angular correlation function using
the ``MegaZ-LRG'' galaxy database, which is a large
photometric-redshift catalogue of luminous red galaxies extracted from
the SDSS imaging data \citep{Collister07}.  They then successfully
model the observations using a HOD model with a ``central'' galaxy
contribution and a ``satellite'' galaxy component.  Noting that
comparison of results are strongly dependent on the overall
normalization of the power spectrum, $\sigma_{8}$, we compare our
effective mass value for the AAOmega LRGs at $z\simeq 0.7$ of $M_{\rm
eff}\simeq 7\times 10^{13} h^{-1} M_\odot$ ($\sigma_{8}=0.85$) to that
of the highest redshift bin by \citet{Blake07} of $0.6 < z < 0.65$ and
find their $M_{\rm eff} = 9.5 \pm 0.7 \times 10^{13} h^{-1} M_\odot$
($\sigma_{8}=0.8$) to be $\sim 30\%$ larger than our effective mass
estimate. However, after further analysis these authors have 
revised their $M_{\rm eff}$ estimates (C. Blake priv. comm) 
and we await comparisons to their new results.

\citet{White07} and \citet{Brown08} have used data from the 
9 deg$^{2}$ Bo$\ddot{\rm{o}}$tes field, 
which has been imaged in the optical and infrared as part of
the NOAO Deep Wide Field Survey \citep[NDWFS;][]{JD99, Brown08},
and by the {\it Spitzer} IRAC Shallow Survey \citep{Eisenhardt04}.
\citet{White07} use the clustering of luminous red galaxies from 
these observations (and $N$-body simulations) to argue that 
about $\frac{1}{3}$ of the most luminous satellite galaxies 
appear to undergo merging or disruption within massive halos 
between $z\simeq0.9$ and 0.5.
\citet{Brown08} report a correlation length of $r_{0}=6.4\pm1.5 \hmpc$ for 
their brightest red galaxy sample, $M_{B} -5 \log h <-21.0$ 
(corresponding to $L>1.6L^{*}$ galaxies), across the redshift range 
$0.6 < z < 0.8$. These authors also calculate the bias for this sample
to be $b=2.15\pm0.08$. 
Thus, although the NDWFS LRGs and AAOmega LRGs have different selections
(e.g. different magnitude and redshift limits), 
evidence from both surveys suggest that redshift $z=0.7$ LRGs 
are highly-biased objects and thus extremely well-suited to LSS studies.

\subsection{LRGs versus ELGs}
\begin{table*}
  \begin{tabular}{lcccccc}
    \hline
    \hline
        {\bf      Scale}    & \multicolumn{2}{c}{\bf ELG}     
                            & \multicolumn{2}{c}{\bf LRG} 
                            & \multicolumn{2}{c}{\bf $V_{\rm eff}$ LRG / $V_{\rm eff}$ ELG}  \\
    \hline
$k/h\, {\rm Mpc^{-1}}$&$P/ h^{-3}\,{\rm Mpc^{3}}$ & $V_{\rm eff}/h^{-3} {\rm Gpc^{3}}$
                      &$P/ h^{-3}\,{\rm Mpc^{3}}$ & $V_{\rm eff}/h^{-3} {\rm Gpc^{3}}$
                      & 167/123  nts. & Equal no. nts. \\
    \hline
0.02  & 6.7$\times 10^{4}$  & 1.1    & $1\times 10^{5}$  & 1.9   &  1.7  &  1.3 \\
0.05  & 2.7$\times 10^{4}$  & 0.82   & $4\times 10^{4}$  & 1.4   &  1.7  &  1.3 \\ 
0.15  & 6.7$\times 10^{4}$  & 0.42   & $1\times 10^{4}$  & 0.61  &  1.5  &  1.1 \\ 
    \hline
    \hline
    \label{tab:LRGs_vs_ELGs_2}
 \end{tabular}
 \caption[A comparison between the effective volumes probed by 
         LRGs vs. ELGs]  
	{A comparison between the effective volumes probed by two
	AAOmega-based BAO Surveys, one using Luminous Red Galaxies
	(LRGs) and one using Emission Line Galaxies (ELGs). We assume a
	factor of 1.5 between the clustering amplitudes of LRGs and
	ELGs.  The second last column is an effective volume ratio for
	\hbox{360 000} LRGs over 3000 deg$^2$ with 70-90\% completeness
	(1.5hr exposures per field) versus \hbox{400 000} ELGs over
	1000 deg$^2$ (1hr exposure) with 80\% completeness both
	assuming 9hr nights. This gives a total observing requirement
	of 167 nights for LRGs and 123 nights for ELGs, implying the
	effective volume ratios given in the sixth column. The last
	column is the effective volume ratio assuming the same number
	of nights for both projects.}
\end{table*}

One of the key questions that the AAOmega LRG Pilot Survey wanted to
address, was whether a ``blue'' or a ``red'' galaxy survey be the more
advantageous when pursuing BAOs at high redshift. In the previous
sections, we have presented the $N(z)$ and clustering amplitudes for
$\bar{z}=0.68$ Luminous Red Galaxies. As such, our `Pilot'
observations suggest, a \ATLAS spectroscopic redshift  survey strategy
to pursue BAOs with AAOmega LRGs might consist of $\approx$1.5 hour
exposures with
\begin{itemize}
\item{ $\simeq 100$ fibres placed on $gri$-selected $i<19.8$ LRGs
  with $z \simeq 0.55$ and}
\item{ $\simeq 260$ fibres placed on $riz$-selected $19.8<i<20.5$ LRGs
  with $z \simeq 0.7$}
\end{itemize}
in order to obtain \hbox{360 000} LRGs over 3000deg$^{2}$  which will
give an $\sim 4 \times$ bigger effective volume than the original SDSS
LRG Survey of 45,000 LRGs  \citep{Eisenstein05}.  
We shall compare this strategy, with an alternate ``Emission Line Galaxy'' (ELG)
survey, in the remainder of this section.

\citet{Glazebrook07} select ``blue'' emission line galaxies (ELGs)
using SDSS and {\it GALEX} Far ultra-violet (FUV) and Near
ultra-violet (NUV) imaging \citep{Martin05},  for the {\it WiggleZ}
BAO Dark Energy Survey. By using the reported $N(z)$ in
\citet[][Figure 2]{Glazebrook07} which has an average redshift of
$z\simeq0.6\pm0.2$ as well as their estimate of the clustering
amplitude,  we can make a comparison with our data.  The clustering
amplitude reported initially by \citet{Glazebrook07} is $s_{0}= 3.81
\pm 0.20 \hmpc$ (their Figure 3). However, it has recently been
suggested that an improved {\it GALEX}  ELG Selection for {\it
WiggleZ} may give a higher ELG clustering amplitude of $r_{0}\approx 6
\hmpc$ (C. Blake priv. comm.) leading to $s_{0}\approx 9 \hmpc$
assuming  $\beta(z\approx0.7)=0.8$ and applying equation 11.  We use
this higher value, along with the appropriate redshift distributions
for ELGs (truncated at redshift $z<0.5$ due to the {\it WiggleZ}
Survey plans to focus on $z>0.5$ galaxies only) and LRGs (from our
Fig. ~\ref{fig:AAOmega_Nofz}) and assuming that bias is scale
independent. 

We can calculate the effective volume surveyed using
\citep[e.g.][]{Tegmark06}:
\begin{equation}
	 V_{\rm eff} = \int \left[
                                  \frac{    n({\bf r}) \, P_{g}(k)}
                                       {1 + n({\bf r}) \, P_{g}(k)}
                             \right]^{2} dV.
\end{equation}
where $n({\bf r})$ is the comoving number density of the sample, (in
units of $h^{3}$ Mpc$^{-3}$) and $P_{g}(k)$ is the value of the galaxy
Power Spectrum at wavenumber $k$ (with units of $h$ Mpc$^{-1}$). For
the LRG Survey we assume $\approx$\hbox{360 000} redshifts are
required with 100 fibres targeted on $i<19.8$, redshift $z\simeq0.55$
2SLAQ LRGs with 90\% completeness, to account for 5\% redshift
incompleteness and 5\% stellar contamination, and 260 fibres on $19.8
< i < 20.5$ $z\simeq0.7$ AAOmega LRGs with 70\% completeness (15\%
redshift incompleteness  and 15\% stellar contamination). For the ELG
Survey, we assume 360 fibres targeted on ELGs, as described above,
with 80\% redshift completeness.  Therefore, we see  that {\it (i)} a
167 night LRG  survey would have $\approx 1.7 \times$ the effective
volume of a 123 night ELG survey as envisaged by Glazebrook et al.
and {\it (ii)} for equal telescope time, an LRG survey will sample
$\approx 1.3 \times$ the effective volume of an ELG Survey (see Table
6).  The above results are approximately in line with those of
\citet{Parkinson07} who present ``Figures of Merit'' (FoM)
calculations to judge the optimality of different survey designs for
future galaxy redshift-based BAO experiments.

\section{Conclusions}
We have reported on the AAOmega-AAT LRG Pilot observing run to
establish the feasibility of a large spectroscopic survey aimed at
detecting BAO and present some of the first results from the new
AAOmega instrument. We have confirmed that AAOmega has a factor of
approximately four in improved throughput in its red ($>5700$\AA) arm as
compared to the old 2dF spectrographs. Utilising this new sensitivity,
we observed  Luminous Red Galaxies (LRGs) selected using single epoch
SDSS $riz$-photometry in 3 fields including the COSMOS field, the
COMBO-17 S11 field and the previously observed 2SLAQ Survey field,
d05. Our main conclusions are:

\begin{itemize}
     \item{We detect 1270 objects in three fields, of which 587 are
       confirmed high-redshift LRGs.  The mean redshift for each
       selection was $\bar{z}=0.578 \pm 0.006$ from the $gri$-band
       selection with $17.5 < i_{\rm deV} < 20.5$, $\bar{z}=0.681 \pm
       0.005$ from the $riz$-band selection with  $19.8 < i_{\rm deV} <
       20.5$ and $\bar{z}=0.698\pm0.015$ from  the $riz$-band selection
       with $19.5 < z < 20.2$.  At $i<20.5$, 84\% redshift completeness
       for LRGs was achieved in 1.67hr exposures in reasonable conditions.}
 
     \item{We have compared our AAOmega spectroscopic redshifts to
       spectroscopic and photometric redshifts obtained by the 2SLAQ
       LRG Survey and COMBO-17 respectively.  We find excellent
       agreement with the 2SLAQ spectroscopic redshifts, but a
       suggestion that there is a  systematic tendency of the
       photometric redshifts to underestimate the spectroscopic
       redshifts by $\overline{\Delta z}=0.026  \pm 0.005$.}
       
     \item{We find that a simple power-law model, for $\wp$,  
       gives a best fit value of $r_{0} =  9.03 \pm 0.93$ for our
       $\bar{z}=0.68$ LRG sample, compared to $r_{0} =  9.80 \pm 0.20$ for
       the $-21.2 < M_{r} < -23.2$ SDSS LRG sample and $r_{0} =  7.30 \pm
       0.34$ for the $\bar{z}=0.55$ 2SLAQ LRG sample. This confirms that
       high-redshift luminous red galaxies are very good large-scale
       structure tracers, similar to their lower redshift counterparts
       \citep{Zehavi05a, Eisenstein05, Ross07}.}

     \item{We also find that, taking into account the large errors on the
           AAOmega LRG $r_{0}$ measurement, there is no inconsistency
           with the simple bias model where the comoving density of LRGs
           are assumed to be constant with redshift.}
       
     \item{Finally, this Pilot project shows that a large-scale AAOmega
       spectroscopic survey of highly biased $z \sim 0.7 $  \hbox{360
       000} LRGs over 3000deg$^{2}$, remains a very promising and
       competitive route in order to measure the baryon acoustic oscillations 
       and use this scale-length to investigate the potential evolution of the
       equation of state  parameter, $w$.}

\end{itemize}

\section*{acknowledgement}

We thank C. Wolf for supplying the COMBO-17 photometric redshift
catalogue data in the S11 field and U. Sawangwit for providing the
Bruzual and Charlot models.  We also thank R. Angulo, C.M. Baugh and
R.M. Bielby for useful discussion.  This work was  supported by a
PPARC PhD Studentship and by National  Science Foundation grant
AST-0607634 (N.P.R.)  We warmly thank all the present and former staff
of the Anglo-Australian Observatory for their work in building and
operating the AAOmega facility. The AAOmega LRG Pilot is based on
observations made with the Anglo-Australian Telescope and with the
SDSS.  Funding for the creation and distribution of the SDSS Archive
has been provided by the Alfred P. Sloan Foundation, the Participating
Institutions, the National Aeronautics and Space Administration, the
National Science Foundation, the U.S. Department of Energy, the
Japanese Monbukagakusho, and the Max Planck Society. The SDSS Web site
is {\tt http://www.sdss.org/}. The SDSS is managed by the
Astrophysical Research Consortium (ARC) for the  Participating
Institutions. The Participating Institutions are The University  of
Chicago, Fermilab, the Institute for Advanced Study, the Japan
Participation Group, The Johns Hopkins University, the Korean
Scientist Group, Los Alamos National Laboratory, the
Max-Planck-Institute for Astronomy (MPIA), the Max-Planck-Institute
for Astrophysics (MPA), New Mexico State University,  University of
Pittsburgh, University of Portsmouth, Princeton University, the United
States Naval Observatory, and the University of Washington.

\setlength{\bibhang}{2.0em}

\appendix

\section{The AAOmega LRG Pilot Data}
In Table A1 we present properties of the first 40 objects from the AAOmega Pilot Catalogue in Right Ascension order. 
The full dataset is published in its entirety in the electronic edition of the {\it Monthly Notices of the Royal 
Astronomical Society}. 

\begin{landscape}
  \begin{table}
    \begin{center}
\tiny
      \begin{tabular}{lcclcrccccccccccccc}
	\hline
	\hline
	Object ID$^{1}$     & $\alpha$   & $\delta$     &
        $r$-fibre   $^{2}$  & T$^{3}$    & X-cor$^{4}$   &
        redshift            & $Q{\rm op}$& Field        &    
       $u$ &  $g$  & $r$ & $i$ & $z$ &
       $u_{\rm err}$ & $g_{\rm err}$ &  
       $r_{\rm err}$ & $i_{\rm err}$ & $z_{\rm err}$ \\
	\hline  
J095618.80+021623.2 & 149.07835 &  2.27312 & 20.14 & 1 &  6.34 & 0.6716 & 3 & cos & 23.6300 & 23.2400 & 21.7760 & 20.5030 & 19.8820 &  1.0310 &  0.2800 &  0.1210 &  0.0730 &  0.1770 \\
J095634.36+015804.9 & 149.14317 &  1.96803 & 19.48 & 1 &  5.51 & 0.5510 & 3 & cos & 22.9990 & 22.0510 & 20.5420 & 19.4830 & 18.8840 &  0.7910 &  0.1420 &  0.0600 &  0.0410 &  0.0910 \\
J095637.67+015254.7 & 149.15698 &  1.88187 & 20.10 & 1 &  3.47 & 0.5686 & 3 & cos & 24.3040 & 22.6720 & 21.4140 & 20.2050 & 19.8920 &  1.2250 &  0.1650 &  0.0890 &  0.0530 &  0.1490 \\
J095639.34+021709.2 & 149.16393 &  2.28590 & 20.27 & 1 &  4.91 & 0.5457 & 3 & cos & 23.4230 & 23.1880 & 21.2570 & 20.2020 & 19.4730 &  1.1030 &  0.3510 &  0.1080 &  0.0720 &  0.1670 \\
J095647.19+022325.2 & 149.19666 &  2.39035 & 19.73 & 1 &  5.57 & 0.5357 & 3 & cos & 25.4570 & 22.6930 & 20.7560 & 19.7260 & 19.7420 &  1.4010 &  0.2240 &  0.0710 &  0.0480 &  0.2130 \\
J095649.24+021838.6 & 149.20519 &  2.31073 & 19.10 & 1 &  5.00 & 0.4806 & 3 & cos & 23.1220 & 21.7150 & 20.0390 & 19.1010 & 18.6840 &  0.8970 &  0.0990 &  0.0400 &  0.0290 &  0.0840 \\
J095649.42+023355.3 & 149.20592 &  2.56537 & 19.93 & 2 &  3.30 & 0.9373 & 1 & cos & 21.5160 & 26.4460 & 21.3550 & 19.9320 & 19.3920 &  0.5760 &  1.6790 &  0.3160 &  0.1550 &  0.3360 \\
J095649.86+020743.4 & 149.20776 &  2.12875 & 20.33 & 2 &  3.06 & 0.9444 & 2 & cos & 22.5640 & 23.1070 & 21.2320 & 20.3290 & 19.4470 &  0.5350 &  0.2970 &  0.0920 &  0.0670 &  0.1380 \\
J095654.01+020225.0 & 149.22507 &  2.04028 & 20.41 & 1 &  6.03 & 0.6589 & 3 & cos & 23.1800 & 23.3520 & 21.9510 & 20.6320 & 20.0930 &  0.6970 &  0.2810 &  0.1320 &  0.0660 &  0.1870 \\
J095702.89+024208.9 & 149.26208 &  2.70250 & 19.78 & 1 &  3.49 & 0.6792 & 4 & cos & 22.2880 & 21.7660 & 20.7220 & 19.7830 & 19.4340 &  0.3290 &  0.0900 &  0.0620 &  0.0470 &  0.1110 \\
J095702.94+014853.3 & 149.26228 &  1.81481 & 20.29 & 1 &  2.98 & 0.7001 & 2 & cos & 24.7150 & 24.6660 & 22.2380 & 20.9450 & 20.6940 &  1.5400 &  0.8110 &  0.2040 &  0.1040 &  0.3320 \\
J095703.84+015133.8 & 149.26600 &  1.85940 & 20.48 & 1 &  3.90 & 0.6355 & 3 & cos & 24.8210 & 22.9440 & 22.1000 & 20.8850 & 20.5140 &  1.2970 &  0.2120 &  0.1840 &  0.1030 &  0.2660 \\
J095705.34+024238.2 & 149.27227 &  2.71061 & 20.46 & 3 &  3.27 & 0.0225 & 2 & cos & 23.4330 & 22.8750 & 22.5620 & 20.7060 & 20.4770 &  1.0860 &  0.3040 &  0.4100 &  0.1370 &  0.3660 \\
J095705.95+021834.5 & 149.27483 &  2.30959 & 20.14 & 1 &  3.47 & 0.8355 & 2 & cos & 26.6360 & 24.5210 & 22.0400 & 20.1440 & 19.9200 &  0.8420 &  1.3890 &  0.3450 &  0.1050 &  0.3750 \\
J095707.64+023955.2 & 149.28184 &  2.66533 & 19.95 & 3 &  7.88 & 0.0010 & 3 & cos & 24.5950 & 22.7660 & 21.6200 & 19.9550 & 19.3470 &  1.5510 &  0.2390 &  0.1550 &  0.0610 &  0.1150 \\
J095708.35+014747.4 & 149.28483 &  1.79652 & 20.16 & 2 &  2.75 & 0.9454 & 2 & cos & 22.0610 & 22.9460 & 21.9640 & 20.1640 & 19.5390 &  1.2060 &  1.0070 &  0.6780 &  0.2210 &  0.5190 \\
J095709.77+020506.6 & 149.29071 &  2.08519 & 20.35 & 3 &  4.97 & 0.0004 & 3 & cos & 22.9240 & 22.7870 & 21.4720 & 20.2040 & 19.7110 &  0.6020 &  0.1890 &  0.0930 &  0.0510 &  0.1410 \\
J095712.04+015554.2 & 149.30018 &  1.93172 & 20.01 & 1 &  3.40 & 0.4818 & 3 & cos & 25.9540 & 22.1380 & 21.0040 & 20.0080 & 19.8880 &  0.7680 &  0.1210 &  0.0780 &  0.0530 &  0.1750 \\
J095712.53+021202.8 & 149.30223 &  2.20079 & 20.26 & 1 &  3.43 & 0.8407 & 3 & cos & 22.0080 & 21.9630 & 21.7840 & 20.5500 & 20.0970 &  0.4650 &  0.1580 &  0.2190 &  0.1200 &  0.3680 \\
J095713.39+015241.0 & 149.30579 &  1.87807 & 19.82 & 1 &  6.31 & 0.6434 & 3 & cos & 24.6310 & 22.6460 & 20.9660 & 19.8160 & 19.2770 &  1.5990 &  0.2100 &  0.0840 &  0.0500 &  0.1140 \\
J095713.89+015210.8 & 149.30789 &  1.86967 & 20.14 & 3 &  2.44 & 0.0449 & 2 & cos & 22.7680 & 22.4500 & 21.5830 & 20.1400 & 19.9110 &  1.0200 &  0.3280 &  0.2630 &  0.1210 &  0.3740 \\
J095719.14+020154.6 & 149.32975 &  2.03184 & 19.85 & 1 &  3.78 & 0.5553 & 3 & cos & 23.0910 & 22.6420 & 20.7870 & 19.8540 & 19.7550 &  0.8480 &  0.2100 &  0.0660 &  0.0470 &  0.1960 \\
J095724.21+013159.5 & 149.35090 &  1.53321 & 19.99 & 1 &  3.06 & 0.5992 & 2 & cos & 22.3180 & 22.9680 & 21.5120 & 19.9900 & 19.5990 &  1.1650 &  0.7730 &  0.3610 &  0.1420 &  0.3850 \\
J095724.98+022905.3 & 149.35408 &  2.48483 & 18.66 & 1 &  5.44 & 0.4814 & 3 & cos & 26.7010 & 21.3450 & 19.5470 & 18.6610 & 18.2400 &  0.7670 &  0.1060 &  0.0340 &  0.0280 &  0.0630 \\
J095728.41+014307.4 & 149.36841 &  1.71873 & 19.39 & 3 &  3.11 & 0.0336 & 1 & cos & 24.3030 & 22.4020 & 20.5450 & 19.3900 & 18.9600 &  2.4410 &  0.2450 &  0.0740 &  0.0430 &  0.1180 \\
J095728.89+021721.2 & 149.37040 &  2.28924 & 20.24 & 3 &  4.21 & 0.0008 & 3 & cos & 25.0320 & 22.8120 & 21.4660 & 20.2770 & 19.7580 &  1.1850 &  0.1700 &  0.0960 &  0.0550 &  0.1390 \\
J095731.70+020327.6 & 149.38210 &  2.05768 & 20.39 & 2 &  2.74 & 0.9269 & 1 & cos & 23.0820 & 22.8770 & 22.2150 & 20.3910 & 19.6020 &  1.0690 &  0.3210 &  0.2880 &  0.0930 &  0.2070 \\
J095733.18+021546.5 & 149.38826 &  2.26293 & 20.26 & 1 &  4.90 & 0.6974 & 3 & cos & 26.0990 & 23.2870 & 21.6230 & 20.4430 & 19.6990 &  0.9610 &  0.4190 &  0.1720 &  0.0980 &  0.2120 \\
J095733.33+013144.5 & 149.38888 &  1.52905 & 20.33 & 1 &  2.52 & 0.6524 & 2 & cos & 23.9740 & 23.1710 & 21.3590 & 20.4520 & 19.5900 &  1.3940 &  0.2940 &  0.1030 &  0.0690 &  0.1170 \\
J095734.28+025024.9 & 149.39286 &  2.84025 & 19.90 & 1 &  6.10 & 0.5540 & 3 & cos & 23.4570 & 22.4010 & 20.8830 & 19.9010 & 19.0370 &  1.0000 &  0.1950 &  0.0840 &  0.0620 &  0.0940 \\
J095737.86+014333.3 & 149.40775 &  1.72593 & 20.44 & 1 &  3.31 & 0.4784 & 1 & cos & 25.9890 & 22.8020 & 21.5670 & 20.2210 & 19.4930 &  1.4960 &  0.3870 &  0.2040 &  0.1020 &  0.2090 \\
J095741.91+020033.7 & 149.42466 &  2.00938 & 19.85 & 1 &  9.80 & 0.6901 & 4 & cos & 22.5380 & 22.3760 & 20.9850 & 19.6760 & 19.2900 &  0.6310 &  0.2390 &  0.1160 &  0.0600 &  0.1570 \\
J095742.07+025028.4 & 149.42533 &  2.84123 & 19.88 & 1 &  6.65 & 0.5144 & 3 & cos & 22.4440 & 22.1470 & 20.7750 & 19.8810 & 19.5570 &  0.3270 &  0.1150 &  0.0570 &  0.0460 &  0.1090 \\
J095742.56+023452.1 & 149.42734 &  2.58114 & 19.42 & 1 & 11.34 & 0.7019 & 5 & cos & 22.5590 & 22.1410 & 20.7490 & 19.4190 & 18.7640 &  0.6400 &  0.1740 &  0.0820 &  0.0430 &  0.0810 \\
J095742.58+024432.1 & 149.42745 &  2.74228 & 20.11 & 3 &  5.51 & 0.0008 & 3 & cos & 22.6610 & 25.3400 & 21.9470 & 20.1110 & 19.6740 &  0.9930 &  2.2170 &  0.4110 &  0.1420 &  0.3160 \\
J095744.56+023835.8 & 149.43567 &  2.64330 & 20.07 & 1 &  6.84 & 0.7358 & 3 & cos & 22.8760 & 23.2550 & 21.5570 & 20.3540 & 19.5620 &  0.6980 &  0.3720 &  0.1420 &  0.0830 &  0.1410 \\
J095744.60+012447.1 & 149.43585 &  1.41309 & 20.11 & 3 &  3.97 & 0.0337 & 1 & cos & 25.3800 & 22.8650 & 21.7010 & 20.1050 & 19.7220 &  3.5850 &  0.6350 &  0.3840 &  0.1410 &  0.3750 \\
J095745.96+014240.1 & 149.44153 &  1.71115 & 20.13 & 3 &  3.05 & 0.0002 & 1 & cos & 24.3480 & 22.5120 & 20.7400 & 19.5390 & 18.7560 &  3.0220 &  0.3360 &  0.1120 &  0.0590 &  0.1220 \\
J095746.79+030025.7 & 149.44499 &  3.00716 & 20.08 & 1 &  3.70 & 0.7130 & 3 & cos & 23.1230 & 23.2530 & 20.6030 & 19.6350 & 18.8280 &  2.1560 &  0.8570 &  0.1320 &  0.0920 &  0.1670 \\
J095748.08+025642.0 & 149.45035 &  2.94503 & 19.85 & 3 &  8.03 & 0.0001 & 3 & cos & 23.1270 & 22.3340 & 21.0780 & 19.4360 & 18.4770 &  0.9510 &  0.1660 &  0.0920 &  0.0350 &  0.0540 \\
\hline
\hline
\label{tab:The_AAOmega_Pilot_Catalogue_Top40}
\end{tabular}
\caption[The first 40 objects from the AAOmega Pilot Catalogue in RA order]
        {The first 40 objects from the AAOmega Pilot Catalogue in RA order.
         The table for the full sample is available online only.
$^1$Using the SDSS nomenclature.
$^2$$r$-band SDSS Fibre Magnitude. 
$^3$Model galaxy template to fit observed spectra. A value of 1 signifies
the ``early-type'' template provided the best-fit, a value of 2 is 
for a $k+a$ Balmer absorption spectrum template and 3 indicates 
the M-star template. 
$^4$Cross-correlation co-efficient between the model and observed galaxy
     spectra.
}
\end{center}
\end{table}
\end{landscape}


\begin{thebibliography}{}
\setlength{\itemindent}{-2.5em}

\bibitem[\protect\citeauthoryear {{Adelman-McCarthy} et~al.,}
{{Adelman-McCarthy} et~al.}{2006}]{Adelman-McCarthy06} 
{Adelman-McCarthy}, J.~K., et al., 2006, \apjs, 162, 38 

\bibitem[\protect\citeauthoryear{{Angulo}, {Baugh}, {Frenk}, \&
{Lacey}}{{Angulo} et~al.}{2008}]{Angulo08}
{Angulo}, R.~E., {Baugh}, C.~M., {Frenk}, C.~S., \& {Lacey}, C.~G.
2008, \mnras, 383, 755 

\bibitem[\protect\citeauthoryear{{Bailey}, {Heald} \& {Croom}}
{{Bailey} et~al.}{2005}]{Bailey05}
{Bailey}, J.~A., {Heald}, R., {Croom}, S,~M., 2005, 
The {\sc 2dfdr} Data Reduction System Users Manual,
AAO, {\tt http://www.aao.gov.au/AAO/2dF/manual.html}

\bibitem[\protect\citeauthoryear{{Baugh}, {Lacey}, {Frenk}, {Granato}, {Silva},
  {Bressan}, {Benson} \& {Cole}}{{Baugh} et~al.}{2005}]{Baugh05}
{Baugh} C.~M.,  {Lacey} C.~G.,  {Frenk} C.~S.,  {Granato} G.~L.,  {Silva} L.,
  {Bressan} A.,  {Benson} A.~J.,    {Cole} S.,  2005, \mnras, 356, 1191

\bibitem[\protect\citeauthoryear{{Bell} et~al.,}{{Bell}
  et~al.}{2004}]{Bell04a}
{Bell} E.~F.,  et~al., 2004, \apjl, 600, L11

\bibitem[\protect\citeauthoryear{{Brown} et~al.,}{{Brown}
  et~al.}{2008}]{Brown08}
{Brown} M.~J.~I.,  et~al., 2008, ApJ submitted

\bibitem[\protect\citeauthoryear{{Blake}, {Collister}, {Lahav}}
{Blake et~al.}{2007}]{Blake07} 
{Blake}, C., {Collister}, A., \& {Lahav}, O.\ 
2007, ArXiv e-prints, 704, arXiv:0704.3377 

\bibitem[\protect\citeauthoryear{{Blake} \& {Glazebrook}}{{Blake} \&
  {Glazebrook}}{2003}]{BlakeGlazebrook03}
{Blake} C. \&  {Glazebrook} K.,  2003, \apj, 594, 665

\bibitem[\protect\citeauthoryear{{Bruzual} \& {Charlot}}{{Bruzual} \&
  {Charlot}}{2003}]{BC03}
{Bruzual} G.,  {Charlot} S.,  2003, \mnras, 344, 1000

\bibitem[\protect\citeauthoryear{{Cannon} et~al.,}{{Cannon}
  et~al.}{2006}]{Cannon06}
{Cannon} R.,  et~al., 2006, \mnras, 372, 425

\bibitem[\protect\citeauthoryear{{Carroll}, {Press} \& {Turner}}{{Carroll}
  et~al.}{1992}]{Carroll92}
{Carroll} S.~M.,  {Press} W.~H.,    {Turner} E.~L.,  1992, \araa, 30, 499

\bibitem[\protect\citeauthoryear{{Cole} et~al.,}{{Cole}  et~al.}{2005}]{Cole05}
{Cole} S.,  et~al., 2005, \mnras, 362, 505

\bibitem[\protect\citeauthoryear{{Colless} et~al.,}{{Colless}
  et~al.}{2001}]{Colless01}
{Colless} M.,  et~al., 2001, \mnras, 328, 1039

\bibitem[\protect\citeauthoryear{{Collister} et al.,}{{Collister}
 et~al.}{2007}]{Collister07} 
{Collister}, A., et al.\ 2007, \mnras, 375, 68 

\bibitem[\protect\citeauthoryear{{Croom} et al.,}{{Croom}
et~al.}{2005}]{Croom05}
Croom, S.~M., et al.\ 2005, \mnras, 356, 415 

\bibitem[\protect\citeauthoryear{{Croom} \& {Shanks}}{{Croom} \&
  {Shanks}}{1996}]{Croom96}
{Croom} S.~M. \&  {Shanks} T.,  1996, \mnras, 281, 893

\bibitem[\protect\citeauthoryear{{da {\^A}ngela} et~al.,}{{da {\^A}ngela}
 et~al.}{2008}]{daAngela08} 
{da {\^A}ngela} J., et al., 2008, \mnras, 383, 565 

\bibitem[\protect\citeauthoryear{{Eisenhardt} et~al.,}{{Eisenhardt}
et~al.}{2004}]{Eisenhardt04}
{Eisenhardt}, P.~R., et al., 2004, \apjs, 154, 48 

\bibitem[\protect\citeauthoryear{{Eisenstein} et~al.,}{{Eisenstein}
  et~al.}{2001}]{Eisenstein01}
{Eisenstein} D.~J.,  et~al., 2001, \aj, 122, 2267

\bibitem[\protect\citeauthoryear{{Eisenstein} et~al.,}{{Eisenstein}
  et~al.}{2003}]{Eisenstein03}
{Eisenstein} D.~J.,  et~al., 2003, \apj, 585, 694

\bibitem[\protect\citeauthoryear{{Eisenstein} et~al.,}{{Eisenstein}
  et~al.}{2005}]{Eisenstein05}
{Eisenstein} D.~J.,  et~al., 2005, \apj, 633, 560

\bibitem[\protect\citeauthoryear{{Eisenstein}, {Seo}, \& {White}}{{Eisenstein}
  et~al.}{2007}]{Eisenstein07}
{Eisenstein}, D.~J., {Seo}, H.-J. \& {White}, M., 2007, \apj, 664, 660 

\bibitem[\protect\citeauthoryear{{Eisenstein} \& {Hu}}{{Eisenstein} \&
  {Hu}}{1998}]{Eisenstein98}
{Eisenstein} D.~J., \& {Hu} W.,  1998, \apj, 496, 605

\bibitem[\protect\citeauthoryear{{Fry}}{{Fry}}{1996}]{Fry96}
{Fry} J.~N.,  1996, \apjl, 461, L65

\bibitem[\protect\citeauthoryear{{Fukugita}, {Ichikawa}, {Gunn}, {Doi},
  {Shimasaku} \& {Schneider}}{{Fukugita} et~al.}{1996}]{Fukugita96}
{Fukugita} M.,  {Ichikawa} T.,  {Gunn} J.~E.,  {Doi} M.,  {Shimasaku} K., \&
  {Schneider} D.~P.,  1996, \aj, 111, 1748

\bibitem[\protect\citeauthoryear{{Glazebrook} et~al.,}{{Glazebrook}
  et~al.}{2007}]{Glazebrook07}
{Glazebrook} K.,  et~al., 2007, astro-ph/0701876

\bibitem[\protect\citeauthoryear{{Gunn} et~al.,}{{Gunn}  et~al.}{2006}]{Gunn06}
{Gunn} J.~E.,  et~al., 2006, \aj, 131, 2332

\bibitem[\protect\citeauthoryear{{Hatton} \& {Cole}}{{Hatton} \&
  {Cole}}{1998}]{Hatton98}
{Hatton} S., \& {Cole} S.,  1998, \mnras, 296, 10

\bibitem[\protect\citeauthoryear{{Hawkins} et~al.,}{{Hawkins}
  et~al.}{2003}]{Hawkins03}
{Hawkins} E.,  et~al., 2003, \mnras, 346, 78

\bibitem[\protect\citeauthoryear{{H{\"u}tsi}}{{H{\"u}tsi}}{2006}]{Hutsi06}
{H{\"u}tsi}, G., 2006, \aap, 449, 891 

\bibitem[\protect\citeauthoryear{{Jannuzi} \& {Dey}}{{Jannuzi} \&
{Dey}}{1999}]{JD99}
{Jannuzi}, B.~T., \& {Dey}, A.\ 1999, 
Photometric Redshifts and the Detection of High Redshift Galaxies, 191, 111 

\bibitem[\protect\citeauthoryear{{Kaiser}}{{Kaiser}}{1987}]{Kaiser87}
{Kaiser} N.,  1987, \mnras, 227, 1

\bibitem[\protect\citeauthoryear{{Kerscher} et~al.}{Kerscher et~al.}{2000}]{Kerscher00} 
Kerscher, M., Szapudi, I., \& Szalay, A.~S., 2000, \apjl, 535, L13 

\bibitem[\protect\citeauthoryear{{Landy} \& {Szalay}}{{Landy} \&
  {Szalay}}{1993}]{LS93}
{Landy} S.~D.,  {Szalay} A.~S.,  1993, \apj, 412, 64

\bibitem[Lawrence et al.(2007)]{2007MNRAS.379.1599L}
{Lawrence}, A., et al., 2007, \mnras, 379, 1599 

\bibitem[\protect\citeauthoryear{{Loh} \& {Strauss}}{{Loh} \&
  {Strauss}}{2006}]{Loh06}
{Loh} Y.-S.,  {Strauss} M.~A.,  2006, \mnras, 366, 373

\bibitem[\protect\citeauthoryear{{Martin} et~al.,}{{Martin}
  et~al.}{2005}]{Martin05}
{Martin} D.~C.,  et~al., 2005, \apjl, 619, L1

\bibitem[\protect\citeauthoryear{{Mart{\'{\i}}nez} \& Saar}{Mart{\'{\i}}nez \& Saar}{2002}]{Martinez02book} 
Mart{\'{\i}}nez, V.~J., \& Saar, E.\ 2002, Statistics of the Galaxy Distribution, Published by Chapman \& Hall/CRC, Boca Raton, ISBN: 1584880848

\bibitem[\protect\citeauthoryear{{Meiksin}, {White}, \& {Peacock}}{{Meiksin}
 et~al.}{1999}]{Meiksin99} 
{Meiksin}, A., {White}, M., \& {Peacock}, J.~A.\ 1999, \mnras, 304, 851 

\bibitem[\protect\citeauthoryear{{Padmanabhan} et~al.,}{{Padmanabhan} 
et~al.}{2007}]{Padmanabhan07}
{Padmanabhan} N., et~al., 2007, \mnras, 378, 852 

\bibitem[\protect\citeauthoryear{{Parkinson}, {Blake}, {Kunz}, {Bassett},
  {Nichol} \& {Glazebrook}}{{Parkinson} et~al.}{2007}]{Parkinson07}
{Parkinson} D.,  {Blake} C.,  {Kunz} M.,  {Bassett} B.~A.,  {Nichol} R.~C.,
\&  {Glazebrook} K.,  2007, \mnras, 377, 185 

\bibitem[\protect\citeauthoryear{{Peebles}}{{Peebles}}{1980}]{Peebles80}
{Peebles} P.~J.~E.,  1980, {The Large-Scale Structure of the Universe}.
Princeton University Press.

\bibitem[\protect\citeauthoryear{{Peebles}}{{Peebles}}{1984}]{Peebles84}
{Peebles} P.~J.~E.,  1984, \apj, 284, 439

\bibitem[\protect\citeauthoryear{{Percial} et~al.,}{{Percival}
 et~al.}{2007a}]{Percival07a} 
{Percival} W.~J., 2007, \apj, 657, 51

\bibitem[\protect\citeauthoryear{{Percial} et~al.,}{{Percival}
et~al.}{2007b}]{Percival07b}
{Percival} W.~J., 2007, \apj, 657, 645 

\bibitem[\protect\citeauthoryear{{Phillipps}, {Fong}, {Fall}, {Ellis} \&
  {MacGillivray}}{{Phillipps} et~al.}{1978}]{Phillipps78}
{Phillipps} S.,  {Fong} R., {Fall} ~S.~M., {Ellis} ~R.~S. \&
{MacGillivray} H.~T., 1978, \mnras, 182, 673

\bibitem[\protect\citeauthoryear{{Phleps}, {Peacock}, {Meisenheimer} \&
  {Wolf}}{{Phleps} et~al.}{2006}]{Phleps06}
{Phleps} S.,  {Peacock} J.~A.,  {Meisenheimer} K., \&  {Wolf} C.,  2006, \aap,
  457, 145

\bibitem[\protect\citeauthoryear{{Press}, {Teukolsky}, {Vetterling} \&
  {Flannery}}{{Press} et~al.}{1992}]{Press92}
{Press} W.~H.,  {Teukolsky} S.~A.,  {Vetterling} W.~T., \&   {Flannery} B.~P.,
  1992, {Numerical Recipes in FORTRAN: The Art of Scientific Computing}.
Cambridge University Press.

\bibitem[\protect\citeauthoryear{{Ross} et~al.,}{{Ross} et~al.}{2007}]{Ross07} 
{Ross} N.~P., et~al., 2007, \mnras, 381, 573 

\bibitem[\protect\citeauthoryear{{Salpeter}}{{Salpeter}}{1955}]{Salpeter55}
{Salpeter} E.~E.,  1955, \apj, 121, 161

\bibitem[\protect\citeauthoryear{{Saunders}, {Rowan-Robinson} \&
  {Lawrence}}{{Saunders} et~al.}{1992}]{Saunders92}
{Saunders} W.,  {Rowan-Robinson} M.,  \& {Lawrence} A.,  1992, \mnras, 258, 134

\bibitem[\protect\citeauthoryear{{Scoccimarro}}{{Scoccimarro}}{2004}]{Scoccima%
rro04}
{Scoccimarro} R.,  2004, \prd, 70, 083007

\bibitem[\protect\citeauthoryear{{Scoville} et~al.,}{{Scoville}  et~al.}
{2007}]{Scoville07} 
{Scoville} N., et~al., 2007, \apjs, 172, 1 

\bibitem[\protect\citeauthoryear{{Seo} \& {Eisenstein}}{{Seo} \&
  {Eisenstein}}{2003}]{Seo03}
{Seo} H.-J., \&  {Eisenstein} D.~J.,  2003, \apj, 598, 720

\bibitem[\protect\citeauthoryear{{Seo} \& {Eisenstein}}{{Seo} \&
  {Eisenstein}}{2005}]{Seo05}
{Seo} H.-J., \& {Eisenstein} D.~J.,  2005, \apj, 633, 575

\bibitem[\protect\citeauthoryear{{Seo} \& {Eisenstein}}{{Seo} \&
  {Eisenstein}}{2007}]{Seo07}
{Seo}, H.-J., \& {Eisenstein}, D.~J.\ 2007, \apj, 665, 14 

\bibitem[\protect\citeauthoryear{{Sharp} et~al.,}{{Sharp}
  et~al.}{2006}]{Sharp06}
{Sharp} R.,  et~al., 2006, \procspie, 6269,  

\bibitem[\protect\citeauthoryear{{Smith} et~al.,}{{Smith} 
et~al.}{2003}]{Smith03} 
Smith, R.~E., et al.\ 2003, \mnras, 341, 1311 

\bibitem[\protect\citeauthoryear{{Stoughton} et~al.,}{{Stoughton}
  et~al.}{2002}]{Stoughton02}
{Stoughton} C.,  et~al., 2002, \aj, 123, 485

\bibitem[\protect\citeauthoryear{{Tegmark} et~al.,}{{Tegmark}
  et~al.}{2006}]{Tegmark06}
{Tegmark} M.,  et~al., 2006, \prd, 74, 123507

\bibitem[\protect\citeauthoryear{{Wake} et~al.,}{{Wake}  et~al.}{2006}]{Wake06}
{Wake} D.~A.,  et~al., 2006, \mnras, 372, 537

\bibitem[\protect\citeauthoryear{{White} et al.}{2007}]{White07}
{White}, M., {Zheng}, Z., {Brown}, M.~J.~I., {Dey}, A., \& {Jannuzi}, B.~T.\ 
2007, \apjl, 655, L69 

\bibitem[\protect\citeauthoryear{{Wolf}, {Meisenheimer}, {Rix}, {Borch}, {Dye}
 \& {Kleinheinrich}}{{Wolf} et~al.}{2003}]{Wolf03} 
{Wolf}, C., {Meisenheimer}, K., {Rix}, H.-W., {Borch}, A., {Dye}, S., 
\& {Kleinheinrich}, M., 2003, \aap, 401, 73 

\bibitem[\protect\citeauthoryear{{Yamamoto}, {Bassett}, {Nichol}, {Suto} \&
  {Yahata}}{{Yamamoto} et~al.}{2006}]{Yamamoto06}
{Yamamoto} K.,  {Bassett} B.~A.,  {Nichol} R.~C.,  {Suto} Y.,    {Yahata} K.,
  2006, \prd, 74, 063525

\bibitem[\protect\citeauthoryear{{York} et~al.,}{{York}  et~al.}{2000}]{York00}
{York} D.~G.,  et~al., 2000, \aj, 120, 1579

\bibitem[\protect\citeauthoryear{{Zehavi} et~al.,}{{Zehavi}
  et~al.}{2005}]{Zehavi05a}
{Zehavi} I.,  et~al., 2005, \apj, 621, 22

\end{thebibliography}
\end{document}